\documentclass[usenatbib]{mn2e}
\usepackage{enumitem}
\usepackage{amsmath}
\usepackage{graphicx}
\usepackage{epsfig}
\usepackage{url}                                 
\usepackage{morefloats}  
\usepackage{amssymb}
\usepackage{rotating}
\usepackage{natbib}
\usepackage{lineno}

\title[Properties of Early-Type Galaxies]{GAMA/H-ATLAS: Linking the Properties of Sub-mm Detected and Undetected Early-Type Galaxies: I. z$\le$0.06 Sample}
\author[N.~K.~Agius et~al.]{\parbox{\textwidth}{N.~K.~Agius,$^{1}$\thanks{E-mail: \texttt{NKAgius@uclan.ac.uk}}
A.~E.~Sansom,$^{1}$
C.~C.~Popescu,$^{1}$
E.~Andrae,$^{2}$
M.~Baes$^{3}$
I.~Baldry$^{4}$
N.~Bourne,$^{5}$
S.~Brough,$^{6}$
C.~J.~R.~Clark,$^{7}$
C.~Conselice,$^{5}$
A.~Cooray,$^{8}$
A.~Dariush,$^{9}$
G.~De~Zotti,$^{10,11}$
S.~P.~Driver,$^{12,13}$
L.~Dunne,$^{14}$
S.~A.~Eales,$^{7}$
C.~Foster,$^{15}$
H.~L.~Gomez,$^{7}$
B.~H\"au\ss ler,$^{5}$
A.~M.~Hopkins,$^{6}$
R.~Hopwood,$^{16,17}$
R.~J.~Ivison,$^{18}$
L.~S.~Kelvin,$^{12,13,19}$
M.~A.~Lara-Lopez,$^{6}$
J.~Liske,$^{20}$
A.~Lopez-Sanchez,$^{6}$
J.~Loveday,$^{21}$
S.~Maddox,$^{14}$
B.~Madore,$^{22}$
S.~Phillipps,$^{23}$
A.~Robotham,$^{12,13}$
K.~Rowlands,$^{5}$
M.~Seibert,$^{22}$
M.~W.~L.~Smith,$^{7}$
P.~Temi,$^{24}$
R.~Tuffs,$^{2}$
E.~Valiante$^{7}$}\vspace{0.4cm}\\
\parbox{\textwidth}{$^{1}$Jeremiah Horrocks Institute, University of Central Lancashire, Preston, PR1 2HE, Lancashire, UK\\
$^{2}$Max Planck Institute for Nuclear Physics (MPIK), Saupfercheckweg 1, 69117 Heidelberg, Germany\\
$^{3}$Sterrenkundig Observatorium, Universiteit Gent, Krijglassn 281 S9, B-9000 Gent, Belgium\\
$^{4}$Astrophysics Research Institute, Liverpool John Moores University, Twelve Quays House, Egerton Wharf Birkenhead, CH41 1LD \\
$^{5}$School of Physics $\&$ Astronomy, The University of Nottingham, University Park Campus, Nottingham NG7 2RD, UK\\
$^{6}$Australian Astronomical Observatory, PO Box 915, North Ryde, NSW 1670, Australia\\
$^{7}$School of Physics $\&$ Astronomy, Cardiff University, The Parade, Cardiff, CF24 3AA, UK\\
$^{8}$Department of Physics $\&$ Astronomy, University of California Irvine, CA 92697,USA\\
$^{9}$Institute of Astronomy, University of Cambridge, Madingley Road, Cambridge, CB3 0HA, UK\\
$^{10}$INAF - Osservatorio Astronomico de Padova, Vicolo Osservatorio 5, I-34136 Trieste, Italy\\
$^{11}$SISSA, Via Bonomea 265, I-34136 Trieste, Italy\\
$^{12}$International Centre for Radio Astronomy Research (ICRAR), University of Western Australia, Crawley, WA 6009, Australia\\
$^{13}$Scottish Universities’ Physics Alliance (SUPA), School of Physics $\&$ Astronomy, University of St Andrews, North Haugh, St Andrews, KY16 9SS\\
$^{14}$Department of Physics $\&$ Astronomy, University of Canterbury, Private Bag 4800, Christchurch, 8140, NZ \\
$^{15}$European Southern Observatory, Alonso de Cordova 3107, Vitacura, Santiago, Chile\\
$^{16}$Physics Department, Imperial College London, South Kensington Campus, London SW7 2AZ, UK\\
$^{17}$Department of Physical Sciences, The Open University, Milton Keynes MK7 6AA, UK\\
$^{18}$UK Astronomy Technology Centre, Science $\&$ Technology Facilities Council, Royal Observatory, Blackford Hill, Edinburgh EH9 3HJ\\
$^{19}$Institut fur Astro-und Teilchenphysik, Universitat Innsbruck, Technikerstrasse 25, 6020 Innsbruck Austria\\
$^{20}$European Southern Observatory, Karl-Schwarzschild-Str. 2, 85748 Garching, Germany\\
$^{21}$Astronomy Centre, University of Sussex, Falmer, Brighton BN1 9QH, UK\\
$^{22}$Observatories of the Carnegie Institution of Washington, 813 Santa Barbara Street, Pasadena, CA 91101, USA\\
$^{23}$Astrophysics Group, School of Physics, University of Bristol, Bristol BS8 1TL, UK\\
$^{24}$Astrophysics Branch, NASA/Ames Research Center, MS 245-6, Moffett Field, CA 94035, USA }}

\begin{document}

\pubyear{2012} \volume{000} \pagerange{\pageref{firstpage}--\pageref{lastpage}} \pubyear{2012}

\maketitle
\label{firstpage}
%\linenumbers
\clearpage

\begin{abstract}

We present two large, nearby (0.013$\le$z$\le$0.06) samples of Early-Type Galaxies (ETGs): a visually classified sample of 220 ETGs, created using source-matched data from the Galaxy and Mass Assembly (GAMA) database with FIR/sub-mm detections from $Herschel$-ATLAS; and a visually classified sample of 551 ETGs which are undetected with $Herschel$-ATLAS. Active galactic nuclei (AGN) are removed from our samples using optical emission line diagnostics. These samples are scrutinised to determine characteristics of sub-mm detected versus undetected ETGs. We find similarities in the stellar mass distributions of the two ETG samples but testing other properties uncovers significant differences. The sub-mm detected sample is shown to have lower concentration and S\'ersic indices than those in the undetected sample - a result which may be linked to the presence of dust in the former. Optical and UV-optical colours are also shown to be much bluer, indicating that the dust is linked with recent star formation. The intrinsic effective radii are on average 1.5 times larger for the sub-mm detected ETGs. Surface densities and groups data from the GAMA database are examined for the two samples, leading to the conclusion that dusty ETGs inhabit sparser environments than non-dusty ETGs in the nearby universe, although environments of the brightest ETGs are shown to differ the least. Modified Planck functions are fit to the H-ATLAS detected PACS and SPIRE fluxes for ETGs with sub-mm flux densities of at least 3$\sigma$ in the 350$\mu$m SPIRE band, giving a resultant mean cold dust temperature of T$_{d}$=22.1K, with a range of 9-30K. The corresponding mean dust mass is 1.8$\times$10$^{7}$M$_{\odot}$, with a range of (0.08-35.0)$\times$10$^{7}$M$_{\odot}$. The dust masses calculated from these fits, normalised by stellar mass, are shown to increase with decreasing stellar mass and bluer colours. Based on visual classifications of elliptical and lenticular, we find similar dust properties for these two early-type morphologies. We conclude that there is a population of elliptical galaxies which exhibit larger dust masses, lower S\'ersic index and bluer colours than the more well-known, massive, red population of ellipticals.

\end{abstract}


\section{Introduction}

%Keep the discussion on morphologies and lead onto visual classification. Then discuss previous survey stuff, previous results from GAMA and H-ATLAS that we want to compare to.

The origin of differences between galaxy morphologies has aroused the interests of researchers for many decades. The original classifications by Hubble laid out a solid foundation for groupings of galaxies. Since then, the Hubble scheme has been reformed multiple times to attempt to explain characteristics associated with galaxies (e.g. \citealp{sandage_2005, cappellari_2011,kormendy_2012}). Although there is still no completely accepted revision to this sequence, galaxies are currently broadly classified into Early-Type Galaxies (ETGs) and Late-Type Galaxies (LTGs).

ETGs form a class of typically quiescent galaxies made up of elliptical (E) and lenticular (S0) galaxies. The ellipticals are pure spheroidal galaxies, while S0s are spheroidal galaxies with a disk, both lacking spiral structure. General properties of this class include a tendency towards redder colours, little or no recent star formation, and mass and light distributions which decline smoothly over large radial ranges \citep{driver_2006}. They can be flattened by rotation or due to velocity anisotropies \citep{illingworth_1977} and their dynamics vary, with the brighter, more massive Es shown to be slow rotators \citep{emsellem_2007}. It can be exceptionally difficult to distinguish between Es and S0s, especially when viewed face-on \citep{vandenbergh_2009}, hence we incorporate both types here.

We intend to further our knowledge of the global characteristics of ETGs in multiple wavebands, with specific emphasis on the properties and origins of ETGs which contain cold dust. This work is achieved by combining data from the Galaxy and Mass Assembly (GAMA) - a wide-area, multi-wavelength spectroscopic and photometric galaxy survey \citep{driver_galaxy_2011} - and the $Herschel$ Astrophysical Terahertz Large Area Survey (H-ATLAS) - a survey that is working towards providing a full dust census of all types of galaxies in the local universe \citep{Eales_2009}. The combination of so many wavebands, from the FUV all the way through to the sub-mm 500$\mu$m band, gives a wealth of information that enables us to study sub-mm detected ETGs and their properties. This is because these wavelengths cover the UV/optical starlight that is absorbed by dust and re-emitted at MIR/FIR/sub-mm wavelengths.

In the past, lack of detections in 21cm line emission (e.g. \citealp{gallagher_1972}) and CO molecular lines (e.g. \citealp{bregman_1992}) in ETG spectra was assumed to indicate a lack of neutral and molecular gas and by implication, cold dust. However, dust lanes and patches were observed in optical images of some elliptical galaxies \citep{sadler_1985, goud_1994}. Developments in FIR and sub-mm based observations have supported these claims, showing that a significant fraction of ETGs contain cold dust within their interstellar medium (ISM; \citealp{knapp_1989,goud_1995}). Some of these were subsequently shown to be spurious detections (see \citealp{bregman_1998}). Studies at different wavelengths have shown this ISM to be complex: radio data led to the detection and mapping of cold neutral gas in some ellipticals (e.g. \citealp{morganti_2006}, and references therein); FIR and sub-mm data revealed significant cold dust masses \citep{leeuw_observations_2004,savoy_scuba_2009}; X-ray observations showed that a substantial number of ellipticals have a hot ($\approx$ few $\times$10$^{6}$K) plasma component (e.g. \citealp{kim_1992,mulchaey_2010,boroson_2011}), particularly in giant ellipticals \citep{kormendy_2009} and older ellipticals \citep{sansom_2006}.

The recent influx of large-scale, multi-wavelength, galaxy surveys provides opportunities to delve further into the dust histories of ETGs. \text{IRAS} was the first infrared telescope to survey dust properties of ETGs, by looking in the MIR/FIR bands at 12$\mu$m, 25$\mu$m, 60$\mu$m and 100$\mu$m \citep{bregman_1998}. FIR emission was detected in 12-17$\%$ of their ETGs. Then, \text{ISO} observed deeper in the FIR, looking at ETGs at 2.5-240$\mu$m wavelengths \citep{temi_2004,xilouris_2004}. The $Spitzer$ space telescope provided infrared observations of some ETGs, mapping their distributions of mid- and far-IR emission \citep{temi_2009}. Optical surveys include the SAURON Integral Field Survey, which enabled a spectroscopic study of ionized gas in ETGs \citep{sarzi_2010}. The ATLAS-3D volume-limited sample of ETGs was used to explore their molecular gas properties \citep{young_2011} and showed that their molecular gas mass was not correlated with galaxy luminosity. A similar result was found for neutral gas \citep{serra_2012}. In the UV the GALEX mission revealed low levels of recent star formation in $\sim$ 30$\%$ of a large sample of ETGs \citep{kaviraj_2007}. Additionally, extended UV disks associated with recent or potential star formation have been identified in small samples of ETGs (see \citealp{moffett_2012}, and references therein). All of these surveys showed that ETGs are not all quiescent, and many contain some cold dust and associated cool gas.

The above mentioned FIR and sub-mm studies are now being superceded by the higher resolution and sensitivity of the $Herschel$ Space Observatory. $Herschel$ SPIRE and PACS data are being used to examine more statistically representative samples of ETGs. Investigations of a volume-limited sample of 62 ETGs, largely in the Virgo Cluster, with the $Herschel$ Reference Survey (HRS), has revealed that some early-types contain low levels of cool dust with temperatures comparable with those of LTGs \citep{smith_HRS}. Nearby galaxies have also been examined in the KINGFISH project with 10 ETGs revealing the possibility of ongoing star formation contributing to the dust heating, as well as heating from the radiation field associated with older stars \citep{skibba_2011}. In the local Universe, H-ATLAS provides the opportunity to work with larger samples. \citet{Rowlands_2011} used the Science Demonstration Phase (SDP) data to show that 42 detected ETGs (5.5$\%$ of luminous ETGs) contain as much dust as some spirals, and \citet{bourne_2012} have used stacking of GAMA sources in H-ATLAS sub-mm imaging to examine the properties of red and blue GAMA galaxies. In this paper and the next (Agius et al, in prep, Part II), we progress beyond these works with much larger samples of H-ATLAS detected ETGs.

One important consideration when working with large samples of ETGs is determining the robustness of the sample morphologies. How can we create a sample which is complete, unbiased and not contaminated by LTGs? Visual inspection is one possibility, but has been shown to be subjective \citep{driver_2006}. The accuracy of eyeballing is also affected by the spatial and flux limitations of the galaxy photometry available from, in this case, the Sloan Digital Sky Survey (SDSS\footnote{\citet{abazajian_2009} - http://www.sdss.org/}). Proxies for morphology can be explored instead, where parameters such as colour \citep{driver_2006,Haines_2008,kaviraj_2010a}, S\'ersic index \citep{leeuw_spatially_2008,blanton_2003,driver_2006,kelvin_2011} and galaxy concentration \citep{blanton_2003,salim_2010} are associated with galaxy type. We investigate such proxies in the future paper. In this paper we make use of an eyeballed sample out to z$\le$0.06 where visual classification is more reliable.

Our investigations explore properties of sub-mm detected versus undetected ETGs. These include the well-known morphology-density relation \citep{dressler_1980}, which shows that ETGs are the dominant galaxy type in high density regions. This has been proven in many studies: for example, \citet{calvi_2011} recently showed that there is a smooth increase in the fraction of ETGs when going from single galaxies, to pairs, to groups. These relations may be explained by hierarchical merging theories, which predict that generations of mergers and interactions have transformed LTGs and built up to the galaxies types we see in the local universe. However there may be some preference for this build-up process to occur on scales of galaxy groups \citep{hoyle_2011}.

If ETGs are formed via merging processes \citep{kormendy_2009} then some may be wet or gas-rich mergers, resulting in residual dust in the merger remnants. Alternatively, could dust produced by old stars reproduce the dust masses currently being seen? The lack of correlations of cold dust mass with host galaxy properties, such as luminosity \citep{knapp_1999,temi_2007} and age \citep{georgakakis_2001}, in ETGs older than $\sim$2 Gyrs suggests that a steady accumulation of dust mass loss from old stars is not likely. The dust may originate externally. We expect most dust to be destroyed via sputtering within $\approx$10$^{7}$ to 10$^{8}$ years \citep{tsai_1995}, if dust is diffusely distributed in a hot X-ray plasma. Therefore the detection of cold dust in ellipticals indicates that this process is not entirely efficient at removing all the dust within this timescale, or the dust is shielded within a cold gas phase, or is being renewed by some other mechanism. %Monolithic collapse models over-predict the metal mass density at high redshifts (sampled by damped Lyman $\alpha$ systems) \citep{madau_1998}. Recent work by \citet{lopez_2012} showed that massive ETGs in the COSMOS survey have undergone an average of 0.89 mergers at z$\le$1, one of many contradictions to monolithic collapse cosmologies.
Thus the history of dust in ETGs is still uncertain.

This paper utilises the GAMA I and H-ATLAS Phase 1 catalogues, described in $\S$\ref{sec:GAMAHATLAS}, and we assume a $\Lambda$CDM cosmology of $\Omega_{M}$=0.3, $\Omega_{\Lambda}$=0.7 and H$_{0}$=100\textit{h}kms$^{-1}$Mpc$^{-1}$ with \textit{h}=0.7. We discuss the parent sample's visual classifications in $\S$\ref{sec:sec2}, as well as additional selection criteria for the two samples of ETGs used in this paper: the optically selected GAMA/H-ATLAS matched sample and the optically selected sub-mm undetected GAMA sample. Diagnostic plots and tests are presented in $\S$\ref{sec:sec3}, including host galaxy properties such as stellar mass, UV luminosities and environments. Dust temperature and dust mass properties are derived and examined in $\S$\ref{sec:sec4}. Conclusions are given in $\S$\ref{sec:conc}.

\section{GAMA and H-ATLAS data}\label{sec:GAMAHATLAS}
%Keep this section in, make the necessary adjustments based on the comments
The GAMA survey is a combined spectroscopic and multi-wavelength imaging campaign designed to study spatial structure in the nearby Universe on scales of 1 kpc to 1 Mpc. For the spectroscopic GAMA I survey, redshifts are measured for galaxies in three regions of 48 deg$^{2}$ each at 9, 12 and 14.5 hours on the celestial equator, for a combined observed area of 144 deg$^{2}$. These regions were selected because of the wealth of multi-wavelength data available from previous optical and NIR surveys. The full survey will measure $\sim$400,000 redshifts and their GAMA I spectroscopic completeness is currently 97$\%$ at SDSS Petrosian magnitude (r$_{pet}) <$19.4 in all three regions \citep{driver_galaxy_2011}. For full details of the GAMA I spectroscopic program, key survey diagnostics and GAMA public and team databases, see \citet{driver_galaxy_2011}.

The GAMA photometry campaign is built up of contributions obtained from multiple facilities and other surveys, including the SDSS for reprocessed optical  \textit{ugriz} photometry \citep{york_2000}, GALEX for FUV and NUV photometry \citep{seibert_2012} and the H-ATLAS survey for FIR/sub-mm photometry. These, in addition to other contributions, provide moderate depth and resolution imaging data across the spectrum. The GAMA input catalogue is described in detail by \citet{baldry_2010} and the spectroscopic tiling of these sources are explained in \citet{robotham_2010}. A full description of the GAMA \textit{ugrizYJHK} photometry pipeline is given in \citet{hill_galaxy_2010}. 

H-ATLAS is the widest area survey with the \textit{Herschel} Space Telescope \citep{Eales_2009}. To survey the largest possible area of sky, the maximum scan rate of 60 arcsec sec$^{-1}$ was used, with both detectors working in Parallel mode to obtain simultaneous observations. These cameras are the Photodetector Array Camera and Spectrometer (PACS), chosen to operate at 100$\mu$m and 160$\mu$m \citep{poglitsch_photodetector_2010}, and the Spectral and Photometric Imaging Receiver (SPIRE), which images at 250$\mu$m, 350$\mu$m and 500$\mu$m \citep{griffin_herschel-spire_2010}. The angular resolutions for these five wavebands are approximately 9$^{\prime\prime}$, 13$^{\prime\prime}$, 18$^{\prime\prime}$, 25$^{\prime\prime}$ and 35$^{\prime\prime}$ full-width half-maximum (FWHM) respectively, with associated 5$\sigma$ point source flux limits of 132, 126, 32, 36 and 45 mJy \citep{rigby_2011}.

The H-ATLAS fields were chosen to maximise the amount of complementary data from other surveys, which results in an overlap with the GAMA equatorial regions described above. This allows source matching between the two catalogues and currently gives large-scale access to $\sim$10,000 galaxies with UV, optical, NIR and FIR/sub-mm data. 

Catalogues for H-ATLAS were created from maps as described in \citet{pascale_2011} and \citet{ibar_2010}. Sources extracted from these maps had to have emission greater than 5$\sigma$ in any of the 3 SPIRE wavebands, described in detail for the SDP in \citet{rigby_2011}. A more extensive description of the Phase 1 dataset will be given by Valiante et al. (in prep). A description of the likelihood-ratio analysis performed to identify robust counterparts to the sub-mm selected sources from the $r$ band matched catalogue can be found in \citet{smith_2011}. This uses both positional and photometric information of both individual sources and of the population in general to determine the reliability of an association between two sources. PACS flux densities are then measured by placing circular apertures at the SPIRE positions. For further descriptions of these methods, refer to \citet{rigby_2011}.

%Find information to comment on confusion and the problem when source matching between cats with such different PSFs.

This paper combines GAMA UV/Optical Phase 1 data and H-ATLAS FIR/sub-mm Phase 1 GAMA-matched sources to create samples of ETGs. All the internal GAMA data are taken from the following Data Management Units (DMUs) available to the GAMA team.

 \begin{itemize}
 \item\noindent TilingCatv16 \\- CATAID, redshift quality \citep{baldry_2010}
 \item\noindent GalexMainv02 \\- FUV and NUV fluxes \citep{seibert_2012}
 \item\noindent S\'ersicCatv07 \\- multiband S\'ersic derived photometry \citep{kelvin_2011}
 \item\noindent StellarMassesv08 \\- rest frame magnitudes, colours and stellar masses derived from optical SED fitting \citep{taylor_2011}
 \item\noindent EnvironmentMeasuresv01 \\- galaxy nearest neighbour surface densities \citep{brough_2012}
 \item\noindent GamaGalaxyGroupsCatv04 \\- galaxy groups and multiplicities \citep{robotham_2011}
 \item\noindent kCorrectionsv02 \\ - GAMA-I multi-band k-corrections 
 \end{itemize}

%Eales et al 2009

%Smith_2011 = SPIRE and PACS map-making processes are described in Pascale et al 2011 and Ibar et al 2010. From these maps a catalogue of sources which are greater than 5sigma in any of the three SPIRE bands was produced using MADX algorithm (maddox et al) and described in detail by Rigby et al 2011 - Need to see if this information is relevant for Phase 1. Smith et al 2011 (the other one) describe the lielihood-ratio analysis performed to identify robust counterparts to the sub-mm selected sources in GALEX-ugrizYJHK matched catalogue, using the SPIRE 250 channel and SDSS r band positions down to a limiting magnitude of SDSS r modelmag=22.4. This uses both positional and photometric information of both individual sources and of the population in general to determine the reliability of an association between two sources.

%Describe how I'm using the most up-to-date data in this paper, etc blah blah

%Is it worth putting the cone plots in here??

%Should have more details about H-ATLAS. What are the properties of the observations? what data do we have access to?  How were the sub-mm/optical IDs done?

\section{Early-type Galaxy Sample Selection}\label{sec:sec2}
%Put in whatever info and references Lee sends me. Discuss the merits of choosing a sample in this way.

To create our nearby sample of ETGs, we took a volume-limited, visually classified sample of GAMA-I galaxies from \citet{kelvin_2013}. This is restricted to the redshift range 0.013$<$z$<$0.06, a Galactic extinction-corrected limit of r$_{pet}<$19.8 and an additional absolute magnitude cut at M$_{r}<$-17.4 to ensure a volume-limited sample. After an additional 19 galaxies were removed due to poor imaging data or bad SDSS photometry, this resulted in a parent dataset of 4,110 galaxies, with a range of galaxy types \citep{kelvin_2013}.

\subsection{Visual Classification}

\begin{figure}
\begin{center}
\hspace*{-0.32in}
 \includegraphics[width=0.52\textwidth]{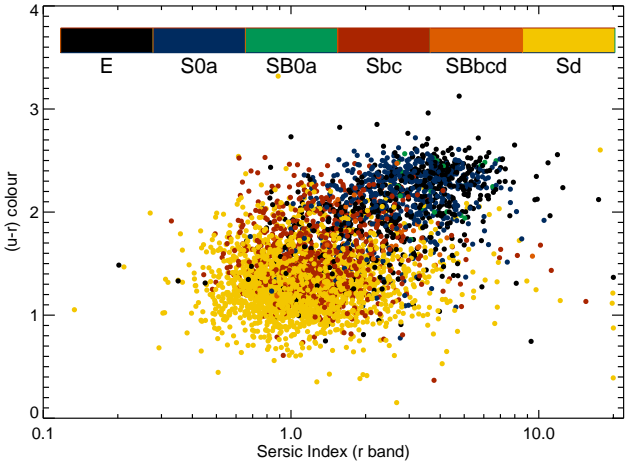} 
 \end{center}
 \caption{Optical colour and S\'ersic index properties of the visually classified parent sample, coloured by galaxy type.}
   \label{fig:ttype}
\end{figure}

The visual classification for this parent sample was carried out and is fully described in \citet{kelvin_2013}. Briefly, they generate three colour postage stamps using UKIDSS H and SDSS $i$ and $g$ bands for red, green and blue colours respectively. These were then classified on a step-by-step basis by three independent people across three tiers. The tiers were defined as follows:

\begin{itemize}
\item[] Is the galaxy spheroid- or disk-dominated?
\item[] Is it a single- or multi-component system?
\item[] Is one component a bar?
\end{itemize}

\noindent These criteria allowed an analysis of the broad-scale morphological properties of the galaxies, independent of colour. Classifications were assigned in cases where at least two of the observers agreed, thereby allowing for objective visual classifications. Spheroid-dominated systems are all classified as ETGs, and then further subdivided into E (single-component) galaxies, S0a (multiple-component) galaxies and SB0a (multiple-component with a bar) galaxies. Note that the S0a grouping includes all sub-catagories  of S0, S0/a and Sa galaxies. Spiral galaxy classifications, which are not the subject of this work, are assigned to systems which are classed as disk-dominated. For example, Sd galaxies are classified as disk-dominated, single-component systems.

The distribution of this parent sample in optical colour and S\'ersic index space is shown in Fig. \ref{fig:ttype}, where S\'ersic indices are extracted from GAMA S\'ersic photometry fits \citep{kelvin_2011} and colours are the restframe colours taken from Stellar Population Synthesis modelling of optical spectral energy distributions (SEDs; \citealp{taylor_2011}). The classifications resulting from the previously described eyeballing are illustrated by colours identified in the colour bar. This plot gives us an indication of the  properties of each galaxy type. It shows that as we go to later types, both colour and S\'ersic index properties decrease on average. This can be interpreted as earlier types tending towards redder colours and less exponential S\'ersic profiles (n$>$1), a result consistent with previous studies such as \citet{peng_2002,blanton_2003,driver_2006,Haines_2008,kaviraj_2010b}.

\subsection{Removal of Active Galactic Nuclei}\label{sec:BPT}

\begin{figure}
\begin{center}
\hspace*{-0.21in}
 \includegraphics[width=0.52\textwidth]{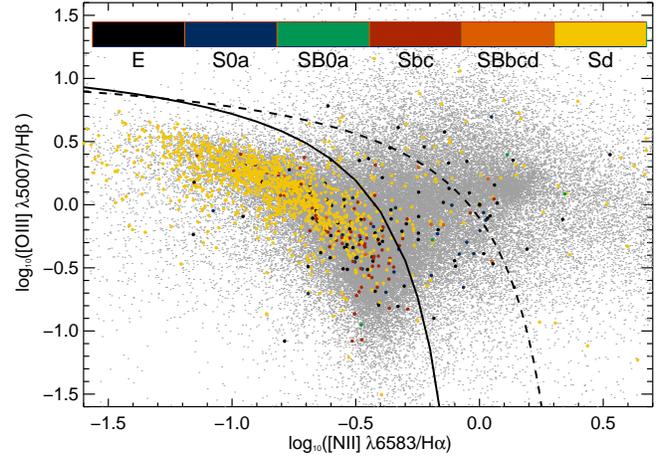} 
 \end{center}
 \caption{This BPT diagram shows the flux ratios of [OIII] $\lambda$5007/H$\beta$ emission lines against [NII] $\lambda$6584/H$\alpha$ emission for all GAMA emission line galaxies (grey points), with the \citet{kauffmann_2003} AGN dividing line shown as the solid black line and the \citet{Kewley_2001} dividing line as the dashed black line. Visually classified galaxies with CATAID-matched emission lines are shown and coloured according to their morphological classification.}
   \label{fig:BPT}
\end{figure}

As the key focus of this work is to compare the properties of two samples of ETGs, it is necessary to make sure the properties obtained accurately represent the host galaxies. In addition, we are not studying the dust heating contribution, or sub-mm emission, from AGN. For these reasons we now discuss our AGN selection method and removal.

	\begin{table*}
   \centering
      \begin{tabular}{l c c c c c c c c c c c c}
      \hline
       &	\multicolumn{2}{c}{\textbf{E}} &	\multicolumn{2}{c}{\textbf{S0a}}	&	\multicolumn{2}{c}{\textbf{SB0a}}	&	\multicolumn{2}{c}{\textbf{Sbc}}	&	\multicolumn{2}{c}{\textbf{SBbcd}}	&	\multicolumn{2}{c}{\textbf{Sd}}\\
       &	\textit{SF}	&	\textit{AGN}	&	\textit{SF}	&	\textit{AGN}	&	\textit{SF}	&	\textit{AGN}	&	\textit{SF}	&	\textit{AGN}	&	\textit{SF}	&	\textit{AGN}	&	\textit{SF}	&	\textit{AGN}	\\
      \hline \hline
	Emission-line galaxies	&	71	&	83	&	20	&	18	&	1	&	3	&	74	&	10	&
	6	&	5	&	1136	&	224\\
	\hline
H-ATLAS detected	&	2	&	2	&	7	&	6	&	0	&	1	&	39	&	5	&	6	&	4	&	27	&	3	\\
	H-ATLAS undetected	&	69	&	81	&	13	&	12	&	1	&	2	&	35	&	5	&	0	&	1	&	1109	&	221	\\	
\hline

   \end{tabular}
   \caption{Galaxies in our parent sample with emission lines are divided into star-forming (SF) and active galactic nuclei emitting (AGN). Numbers shown here are divided into H-ATLAS 250$\mu$m flux detected (102 galaxies) and those which are undetected by H-ATLAS (1549 galaxies). They are additionally separated into their morphological classifications.}
   \label{tab:AGNnums}
\end{table*}

\citet{Foster_2012} use the Gas and Absorption Line Fitting (GANDALF; \citealp{sarzi_2006}) algorithm to measure stellar emission lines from flux calibrated GAMA spectra. They define galaxies with significant emission lines as those with 3$\sigma$ levels in H$\alpha$, H$\beta$ and [NII]$\lambda$6584 lines. With their results, we produce a BPT diagram \citep{baldwin_1981} for GAMA galaxies with these lines (see Fig. \ref{fig:BPT}). Emission-line galaxies in our parent sample are over-plotted to indicate their distribution and are again coloured by eyeballed morphology. Based on the prescription by \citet{kauffmann_2003} (Eq. 1), we show the dividing line between star-forming galaxies and AGN-dominated galaxies, in the form:

\begin{equation}
\log\left(\frac{[OIII]}{H\beta}\right) = \frac{0.6}{\log\left(\frac{[NII]}{H\alpha}\right)-0.05} + 1.3
\label{eq:kauff}
\end{equation}
			
This line identifies the upper limit for emission line measurements which are representative of pure star formation in the BPT plot. Fig. \ref{fig:BPT} also shows the \citet{Kewley_2001} upper limit line (Eq. \ref{eq:kewley}) for extreme starburst limits.

\begin{equation}
\log\left(\frac{[OIII]}{H\beta}\right) = \frac{0.61}{\log\left(\frac{[NII]}{H\alpha}\right)-0.47} + 1.19
\label{eq:kewley}
\end{equation}

\noindent Note that this latter relationship is not our AGN identifier and is only shown for comparison as a soft limit. All galaxies above and to the right of the Kauffmann dividing line are considered to have optical AGN emission.

 When matching our parent sample to the GANDALF emission lines catalogue, we get 1651 (40$\%$) matches for emission lines. From this selection, 343 (21$\%$) galaxies are dominated by AGN and 1308 (79$\%$) are star-forming. Lack of significant emission lines is the chief cause of unmatched galaxies. The colour selection in Fig. \ref{fig:BPT} indicates how these are divided by morphology and we show the numerical results in Table \ref{tab:AGNnums}. The first row in Table \ref{tab:AGNnums} shows that early-types (E, S0a and SB0a) have larger proportions of their emission-line galaxies identified as AGN ($\sim$59$\%$) than later-types ($\sim$26$\%$). However, both Table \ref{tab:AGNnums} and Fig. \ref{fig:BPT} show that late-types preferentially occupy the BPT diagram and for our parent sample, emission from these late-types is most likely powered by young stars rather than AGN.
 
 For the next stage of our sampling, we remove all galaxies with optically identified AGN emission from our parent sample, leaving us with 3767 galaxies to pick ETGs from. The following section describes how we use those galaxies classified as E or S0a by \citet{kelvin_2013} to form two samples which are the subject of this work.

\subsection{H-ATLAS Detected and Undetected Samples}\label{sec:samples}

%Discuss the final sample. Show a couple of colour images for example galaxies in each sample. Explain that they're volume limited. Show the rmag vs redshift plot.
We chose to create two samples out of the remaining visually classified ETGs: the optically selected, sub-mm detected, ETG Sample ($SubS$), and the optically selected, sub-mm undetected, ETG Sample ($OptS$). $SubS$ is the eyeballed sample of E (elliptical) and S0a (lenticular/early Sa spiral) galaxies which have positional matched detections to within 1$^{\prime\prime}$ in both GAMA and H-ATLAS. $OptS$ is the eyeballed sample of GAMA ETGs which do not have H-ATLAS detections (5$\sigma$ in any SPIRE waveband). As described, our samples contain Sa galaxies, but we make the assumption that they are not numerous and will not significantly skew our results. This is likely due to additional imposed selection criteria (described below), together with published Sa population percentages ($\sim$18$\%$ of ETGs; \citealp{Nair_2010}, their Table 3). 

\begin{figure}
\begin{center}$
\begin{array}{ccc}
  \includegraphics[width=0.15\textwidth]{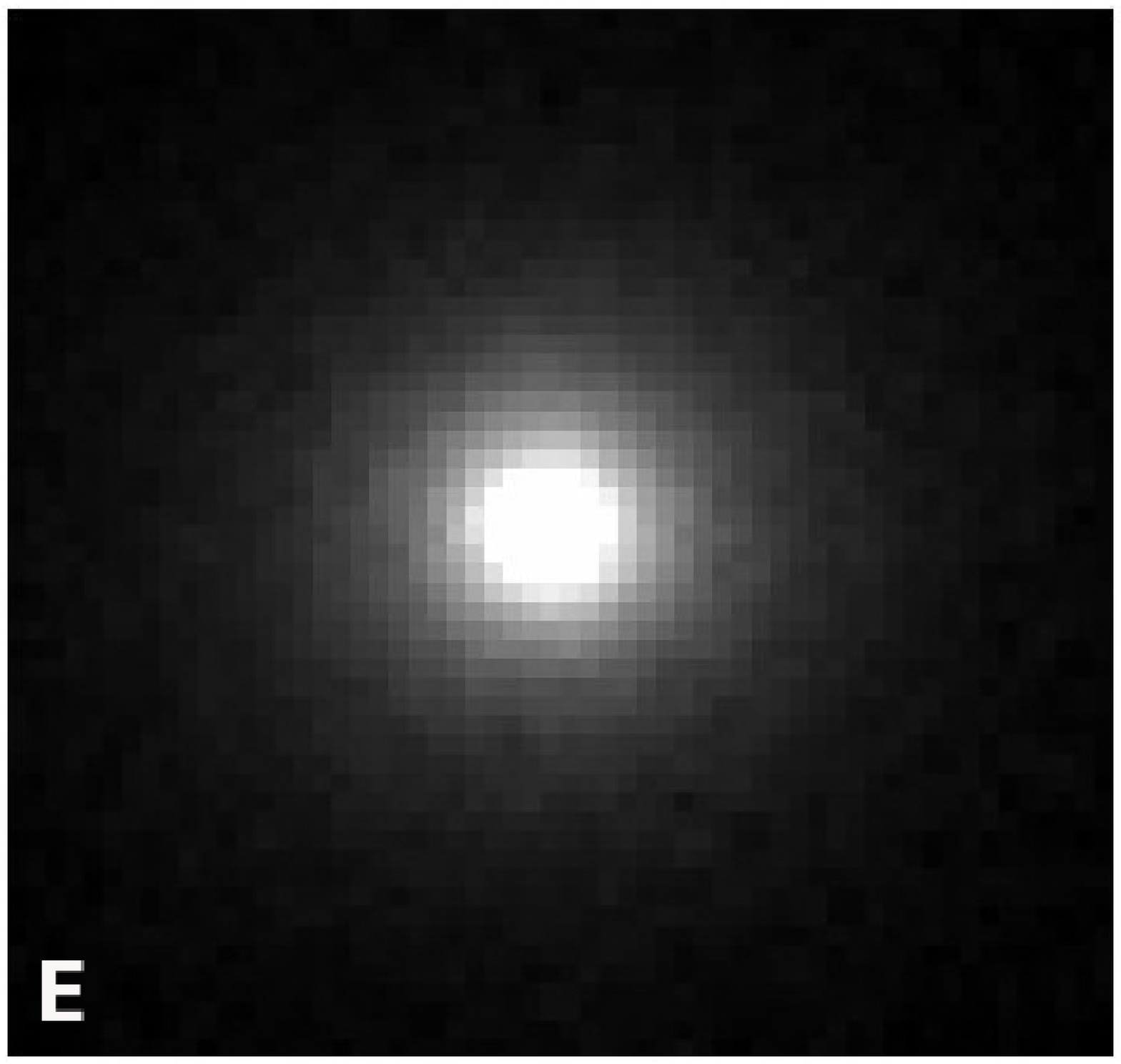} &
   \includegraphics[width=0.15\textwidth]{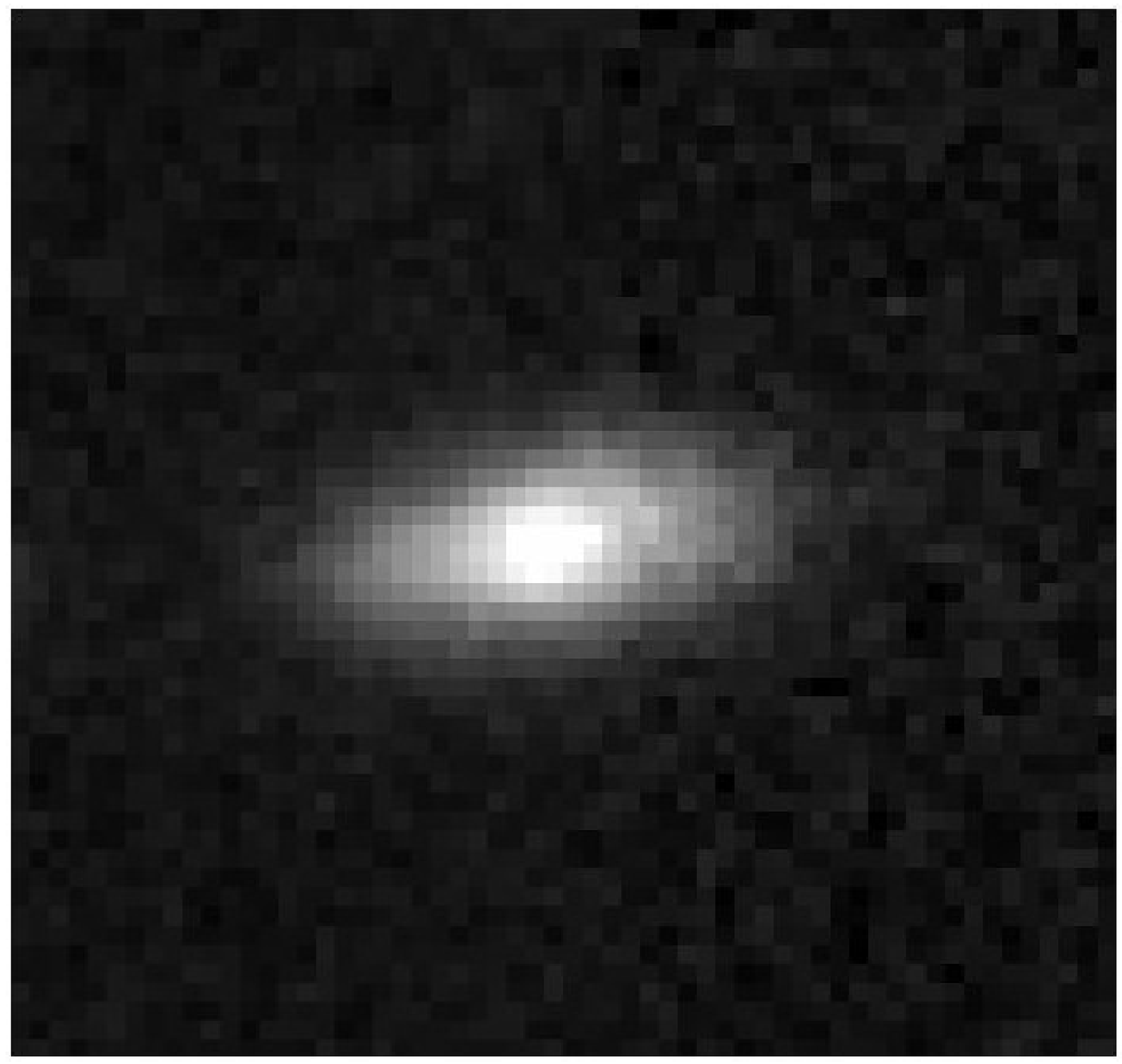} &
    \includegraphics[width=0.15\textwidth]{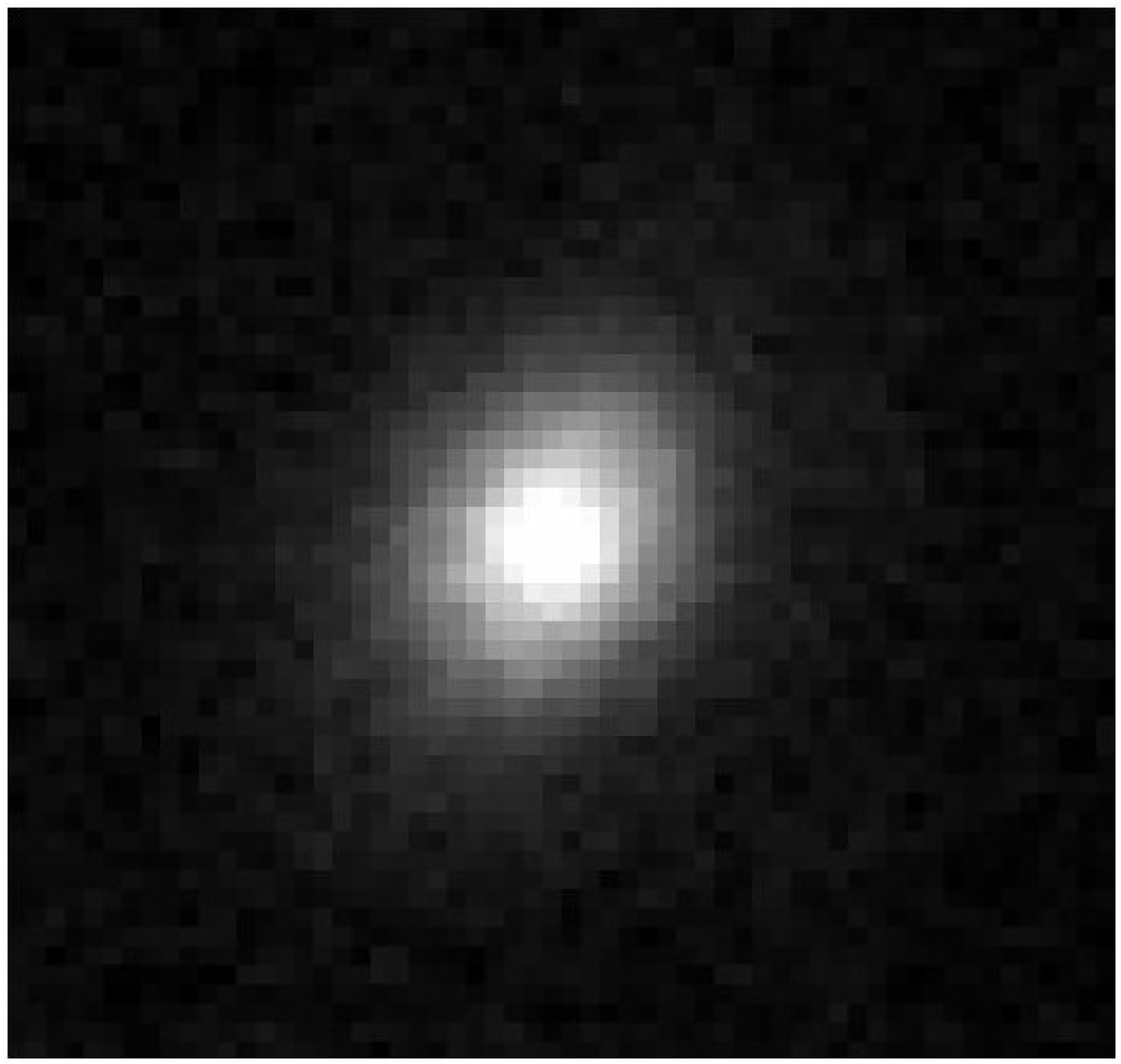} \\
      \includegraphics[width=0.15\textwidth]{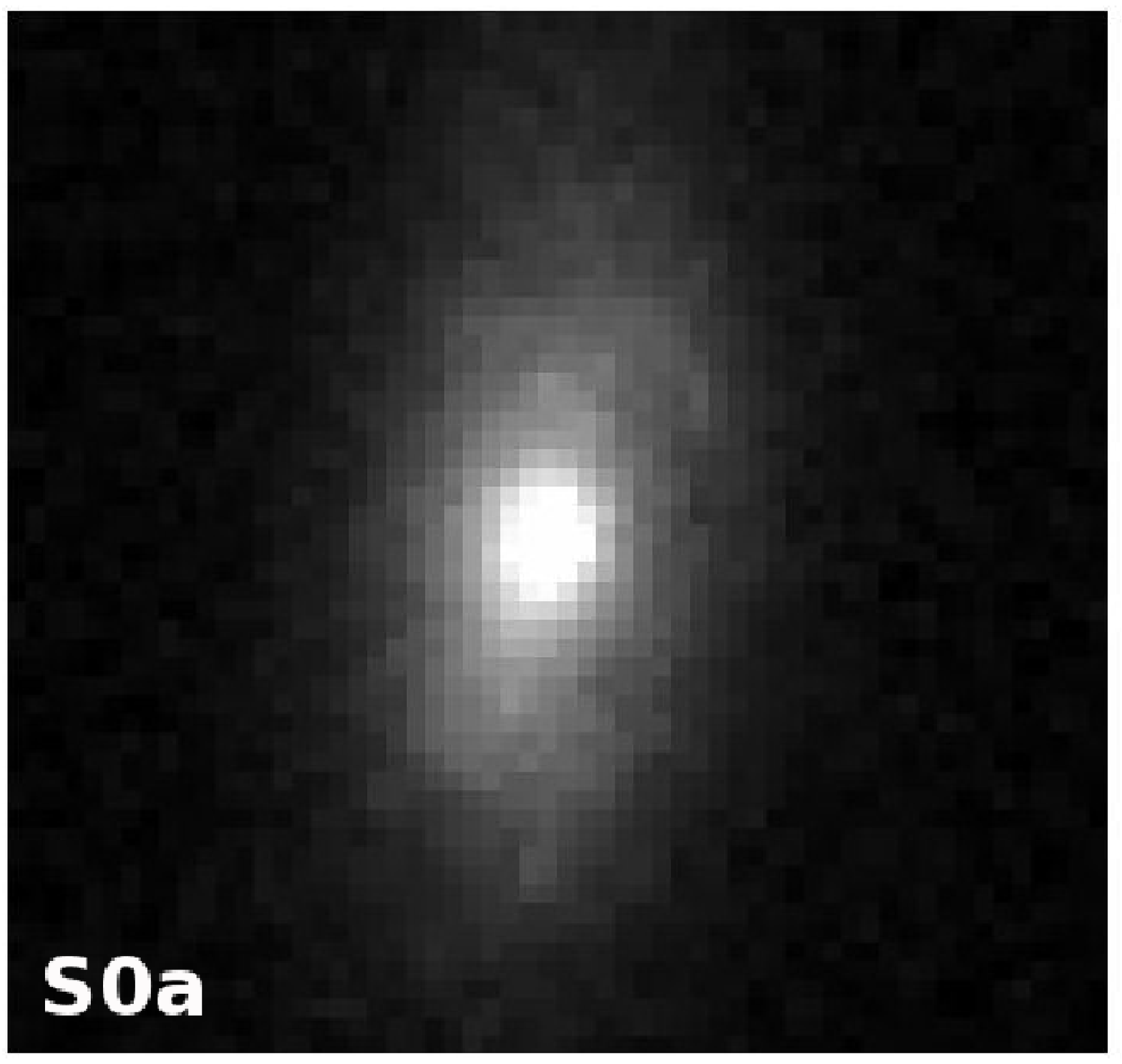} &
       \includegraphics[width=0.15\textwidth]{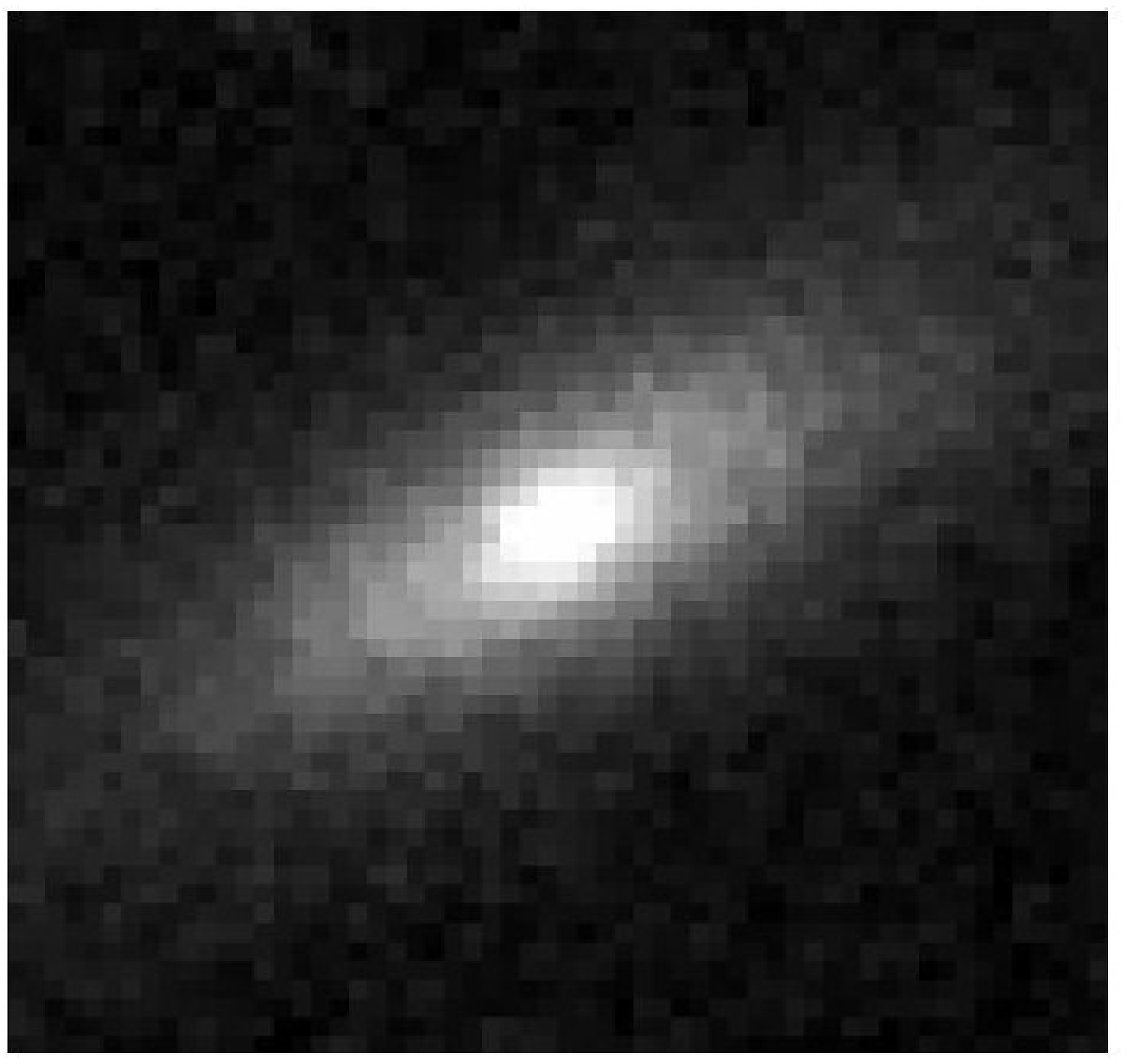} &
       \includegraphics[width=0.15\textwidth]{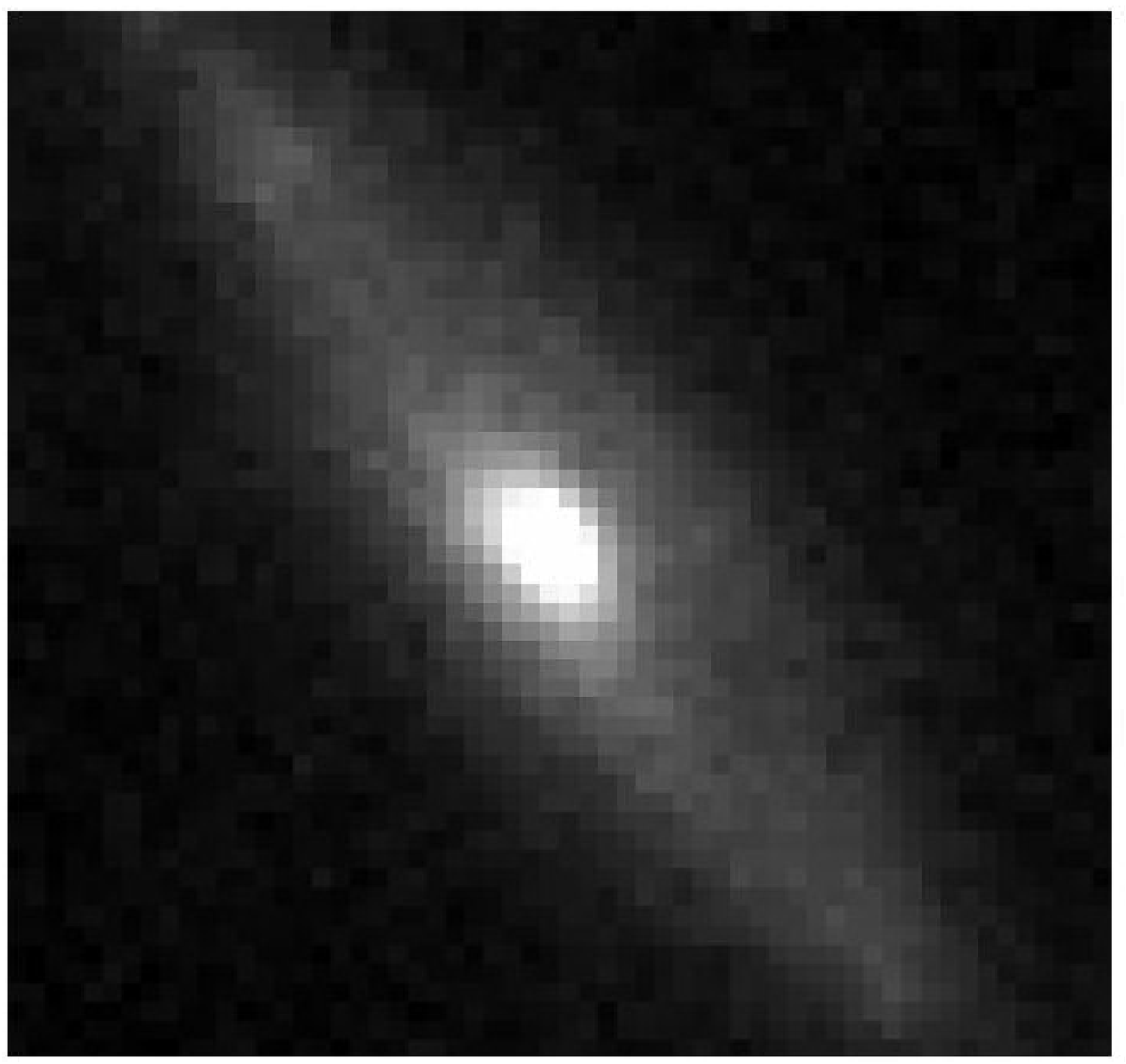} \\  

   \end{array}$
 \end{center}
 \caption{Example images of sub-mm detected galaxies with E classifications (top row) and S0a classifications (bottom row). The images are 20$^{\prime\prime}$ SDSS $g$ band cutouts. }
 %The images are 100$^{\prime\prime}$ cutouts, on a 0.198 $^{\prime\prime}$/pixel scale.

   \label{fig:images}
\end{figure}

We also imposed additional selection criteria to form these samples. This includes setting an ellipticity cut of $(1-(b/a))\le 0.7$ to remove any edge-on disk galaxies. Additionally we ensured that all our sample ETGs had an effective radius of at least the seeing FWHM (1.2$^{\prime\prime}$) to minimise any uncertainty in parameters related to seeing. These criteria removed 32 and 62 galaxies from the $SubS$ and $OptS$, respectively. As described in $\S$\ref{sec:BPT}, we removed galaxies characterised by optical AGN emission, showing these numbers in rows 2 and 3 of Table \ref{tab:AGNnums} for H-ATLAS detected and undetected galaxies respectively. For the H-ATLAS detections, we also removed 15 galaxies which had low reliabilities and are therefore considered likely to be false counterparts. Finally, we ran an independent visual check on these galaxies to remove any with obvious spiral arms. This was a necessary additional criterion as spiral arms were not accounted for in the eyeballing by \citet{kelvin_2013}, and 22 galaxies were removed in this process. Our resultant samples included 229 ETGs (33$\%$ E and 67$\%$ S0a) in the $SubS$ and 551 ETGs (54$\%$ E and 46$\%$ S0a) in the $OptS$. Therefore the sub-mm detected ETGs are more dominated by lenticulars. Examples of $SubS$ ETGs are shown in Fig. \ref{fig:images}.

\begin{figure}
\begin{center}
 \hspace*{-0.25in}
   \includegraphics[width=0.52\textwidth]{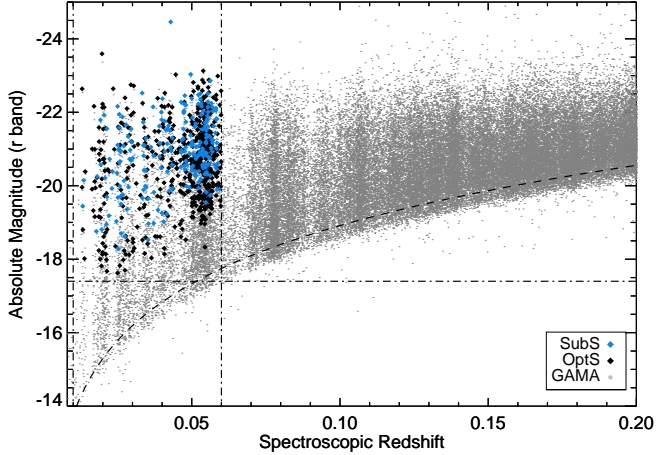} 
  \end{center}
 \caption{The distribution of all GAMA detections (grey points) plotted as a function of redshift up to z=0.2. The dashed line shows the GAMA r band spectroscopic completeness limit of m$_{r}$=19.4. The vertical dashed lines illustrate the redshift cutoffs for our early-type samples and the horizontal dashed line indicates the absolute magnitude cutoff. The sub-mm detected sample is shown as blue diamonds and the undetected sample as black diamonds.}
   \label{fig:red1}
\end{figure}

 \begin{figure*}
\begin{center}
\hspace*{-1cm}
 \includegraphics[width= 0.38\textwidth]{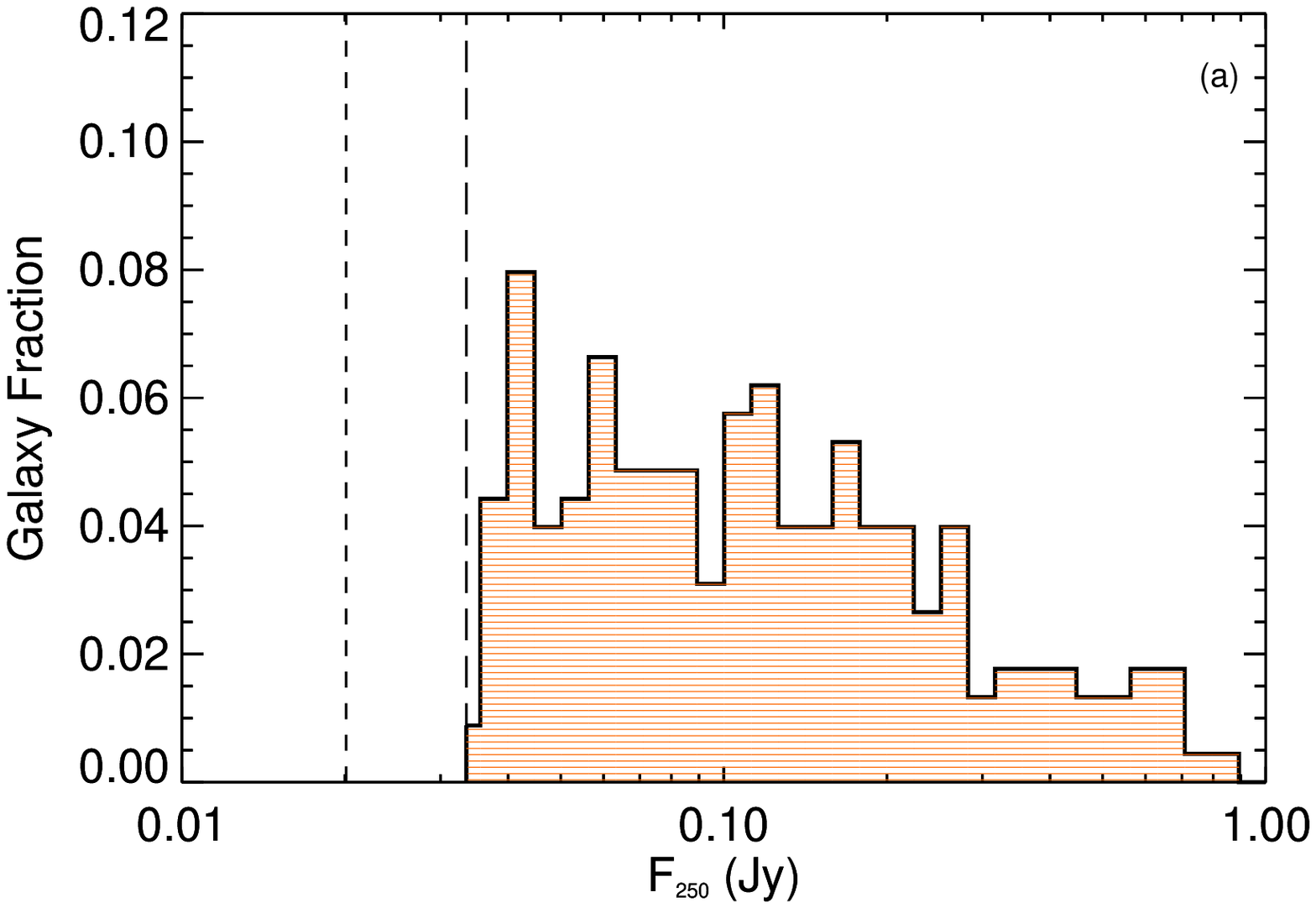} 
 \hspace*{-1cm}
  \includegraphics[width= 0.38\textwidth]{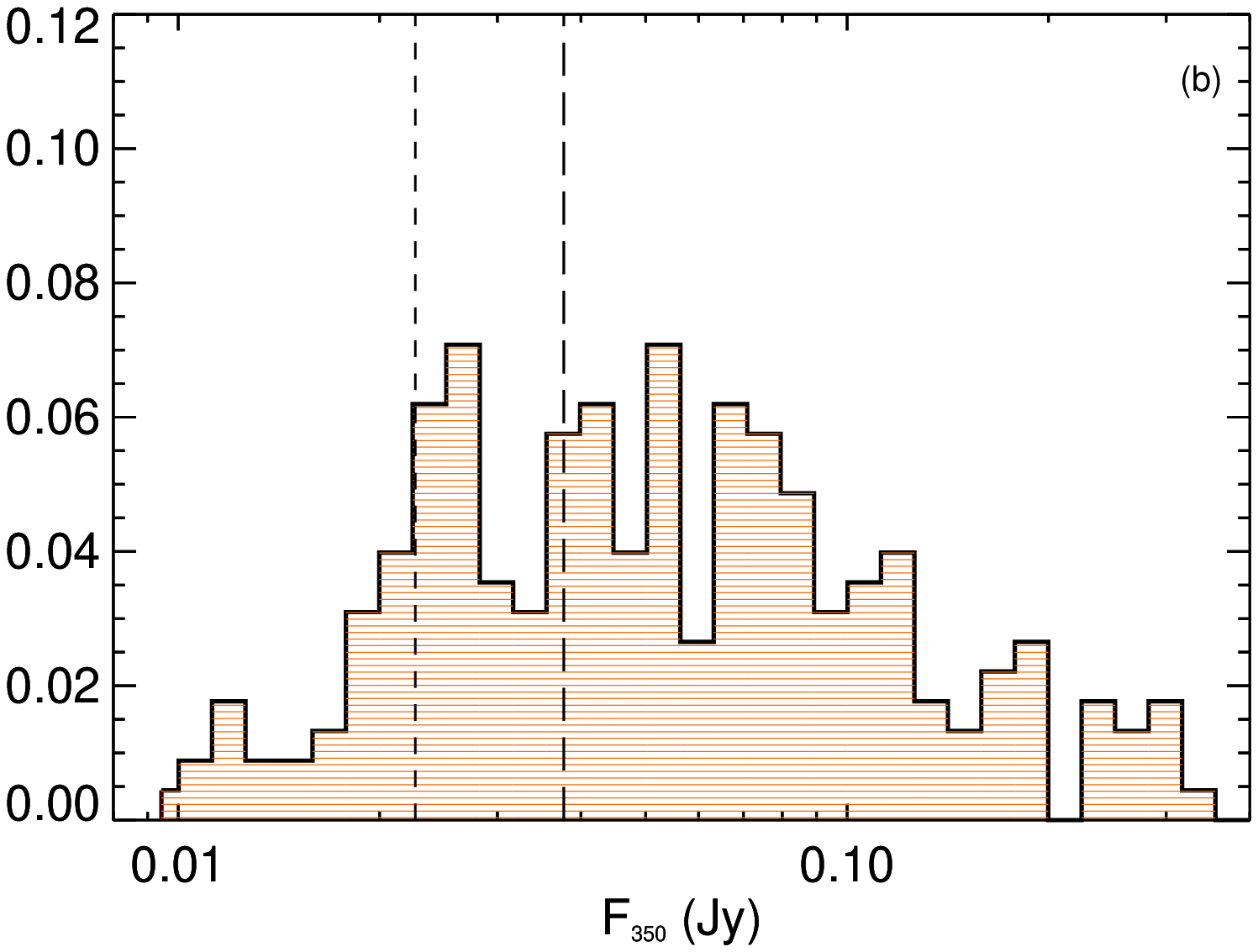} 
  \hspace*{-1cm}
   \includegraphics[width= 0.38\textwidth]{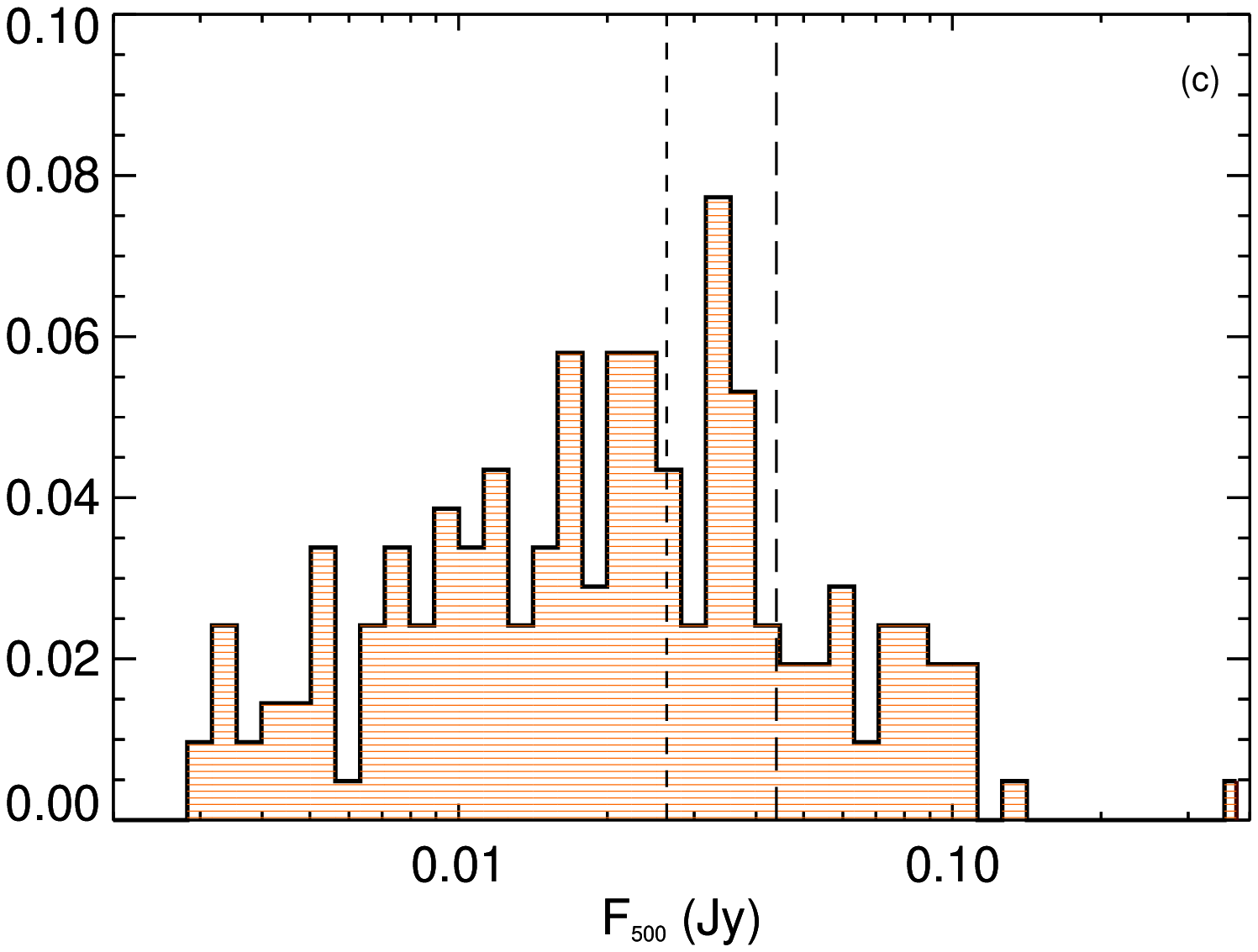} 
 \end{center}
 \caption{The orange histograms show (a) 250$\mu$m, (b) 350$\mu$m and (c) 500$\mu$m H-ATLAS flux values for our sample of optically-selected, early-type galaxies with 5$\sigma$ detections in at least one band. The short dashed line gives the 3$\sigma$ flux level, and the long dashed line the 5$\sigma$ flux level \citep{Eales_2009}. The distribution of the $SubS$ is above these levels in the 250$\mu$m regime, due to the selection criteria, but $\sim$35$\%$ of 350$\mu$m fluxes are below the 5$\sigma$ level and most of the distribution lies below the 5$\sigma$ flux level for the 500$\mu$m band. }
   \label{fig:fluxes}
\end{figure*}

	\subsection{Selection Effects and Completeness}\label{sec:SelEff}
	
	Both $SubS$ and $OptS$ are affected by selection effects and completeness issues. In this section we will examine these and try to assess how they will affect our comparisons.
	
	First of all, both samples are affected by a lack of GAMA catalogue ID (CATAID) completeness. The 229 and 551 ETGs in the $SubS$ and $OptS$ respectively all have GAMA CATAIDs and redshifts with normalised redshift qualities nQ$\ge$3 (good for science). However, due to the failure in extraction of derived parameters for a small subset of galaxies in the GAMA database, our samples do not 100$\%$ match CATAIDs in all GAMA internal catalogues used in this work. In particular, NUV magnitudes are not complete for both samples. We explore the level of this incompleteness in $\S$\ref{sec:UV}.
	 
	 Fig. \ref{fig:red1} shows the GAMA absolute r band magnitude of the two samples, as a function of redshift. The plot marks the r band spectroscopic completeness limit of r$_{pet}=$19.4 in all three fields as the dashed line \citep{driver_galaxy_2011}. Note that there are galaxies plotted below this limit which will be as faint as r$_{pet}$=19.8: the magnitude limit for the ongoing GAMA-II data collection. This plot also highlights the absolute magnitude cutoff used here to form a volume-limited sample (down to $\textrm{M}_{r}= -17.4 \textrm{mag}$., i.e. slightly brighter than the Small Magellanic Cloud). Less luminous galaxies are removed from our samples in this paper. A volume-limited sample going out to higher redshift would require a brighter magnitude cut, an issue which will be dealt with in Paper II of this series.
	 
	 This plot shows an apparent lack of completeness at higher redshift for our samples. The parent sample is complete even in this regime and therefore this effect is a result of the methods used to create the two samples. We investigate this phenomenon on a galaxy-by-galaxy basis and find the removal of AGN and galaxies with low effective radius are the main cause of this incompleteness. However, this effect is true for both samples, indicating that comparative measures between the two are unlikely to be biased.
	%Put in a discussion of completeness levels in the relevant GAMA bands
	
	%Discuss Fig 5 since it shows the absolute magnitude limits reached at different z, due to the flux limited nature of the GAMA survey.
		
	We now discuss completeness issues specific to \textit{SubS}. These sources were selected based on H-ATLAS SPIRE detections greater than 5$\sigma$ in any waveband. As shown in Fig. \ref{fig:fluxes}(a), all these 5$\sigma$ detections are in the 250$\mu$m waveband. A study in \citet{rigby_2011} shows H-ATLAS SDP as having greater than 80$\%$ catalogue number density completeness. This missing $\sim$20$\%$  is due to a number of undetected faint sources indicated by random noise fluctuations in the simulated maps, or because of source blending. This is likely to be larger for the Phase 1 data due to its deeper region. \citet{dunne_2010} also shows a slight breakdown in ID completeness, as they cannot guarantee that all detections which are given SDSS ID counterparts have been correctly identified. This is due to positional uncertainties, close secondaries and the random probability of finding a background source within 10$^{\prime\prime}$. 
	
	Based on recommendations by \citet{smith_2011}, all of our sub-mm galaxies have a reliability R$>$0.8, where reliability is the likelihood that each object is the correct counterpart out of all the counterparts within a search radius. This ensures that the contamination rate is kept to a minimum and that the SDSS r-band source is linked to the FIR emission. Based on this value, we are able to estimate the likely number of false IDs in the \textit{SubS}. This is done using Eq. 12 from \citet{smith_2011}:
	
	\begin{equation}
	N(false) = \sum_{R>0.8}(1-R)
	\end{equation}

	This results in $\sim$3 galaxies (1$\%$) in our sample having a possibly false ID. We judge this fraction low enough to not be a major problem for the rest of this work. 

	We test the galaxies in \textit{SubS} to check for the likelihood of contamination from strong lensing sources. We use two methods to check for lenses. The first method is a study by \citet{negrello_2010}, where galaxies with SPIRE 500$\mu$m emission $\ge$ 100 mJy are likely to be strong lenses. Seven of the galaxies in the \textit{SubS} fulfill this criterion. The second method is that of \citet{Gonzalez-Nuevo:2012ys}, where galaxies fulfilling all the following criteria have a 50$\%$ likelihood of being strong lenses.
	
	\begin{itemize}
	\item[] SPIRE 250 $\mu$m flux emission $\ge$ 35mJy\\
	\item[] SPIRE 350 $\mu$m flux emission $\ge$ 85mJy\\
	\item[] SPIRE 350/250 $\mu$m ratio $\ge$ 0.6\\
	\item[] SPIRE 500/350 $\mu$m ratio $\ge$ 0.4\\
	\end{itemize}

	\noindent Based on this second set of criteria, we find two ETGs in the \textit{SubS} are 50$\%$ likely to be strong lenses. However, these two methods do not pick out the same galaxies as being possible strong lenses. Altogether, these tests indicate low ($\sim$4$\%$) levels of contamination from lensing, but we remove these nine galaxies from the $SubS$ as we want to avoid including any possible strong lenses or background contaminants in our sample. This reduces our $SubS$ to 220 ETGs.

	The second and third plots in Fig. \ref{fig:fluxes} indicate that in the 350$\mu$m band, 65$\%$ (142 detections) of the galaxies are above 5$\sigma$ and only 14$\%$ (31 detections) are above this limit in the 500$\mu$m waveband. This causes some problems, as will be discussed later in $\S$\ref{sec:sec4}, with fitting SEDs to the data to obtain dust-related parameters.

% 	Another question about the \textit{SubS} is how much it is limited by the optical. It is of interest to check if there are any sub-mm detected galaxies which are unseen by GAMA or the SDSS. By comparing the SDSS IDs and GAMA Galaxy Catalogue IDs with the 250$\mu$m 5$\sigma$ detections we discovered that there are 35,720 (59$\%$) of H-ATLAS sources which are not recorded in these two optical databases. Although these detections may not significantly affect this work (as they are unlikely to be brighter than M$_{r}$=-17.4, nearby ETGs), they will be of interest to future work. Some will be lensed sources. For our detected \textit{SubS} ETGs, statistical separation of their intrinsic sub-mm emission  from components of sub-mm emission arising from combined effects of blending and lensing is work currently under progress.

	In subsequent sections everything is examined on a comparative basis between the purely optical and optical+sub-mm samples whenever possible. This comparative approach allows us to investigate the relative behaviours of sub-mm detected versus undetected ETGs in the nearby universe.

\section{Sub-mm Detected vs Undetected Diagnostics}\label{sec:sec3}

\begin{figure*}
\begin{center}
 \includegraphics[width=\textwidth]{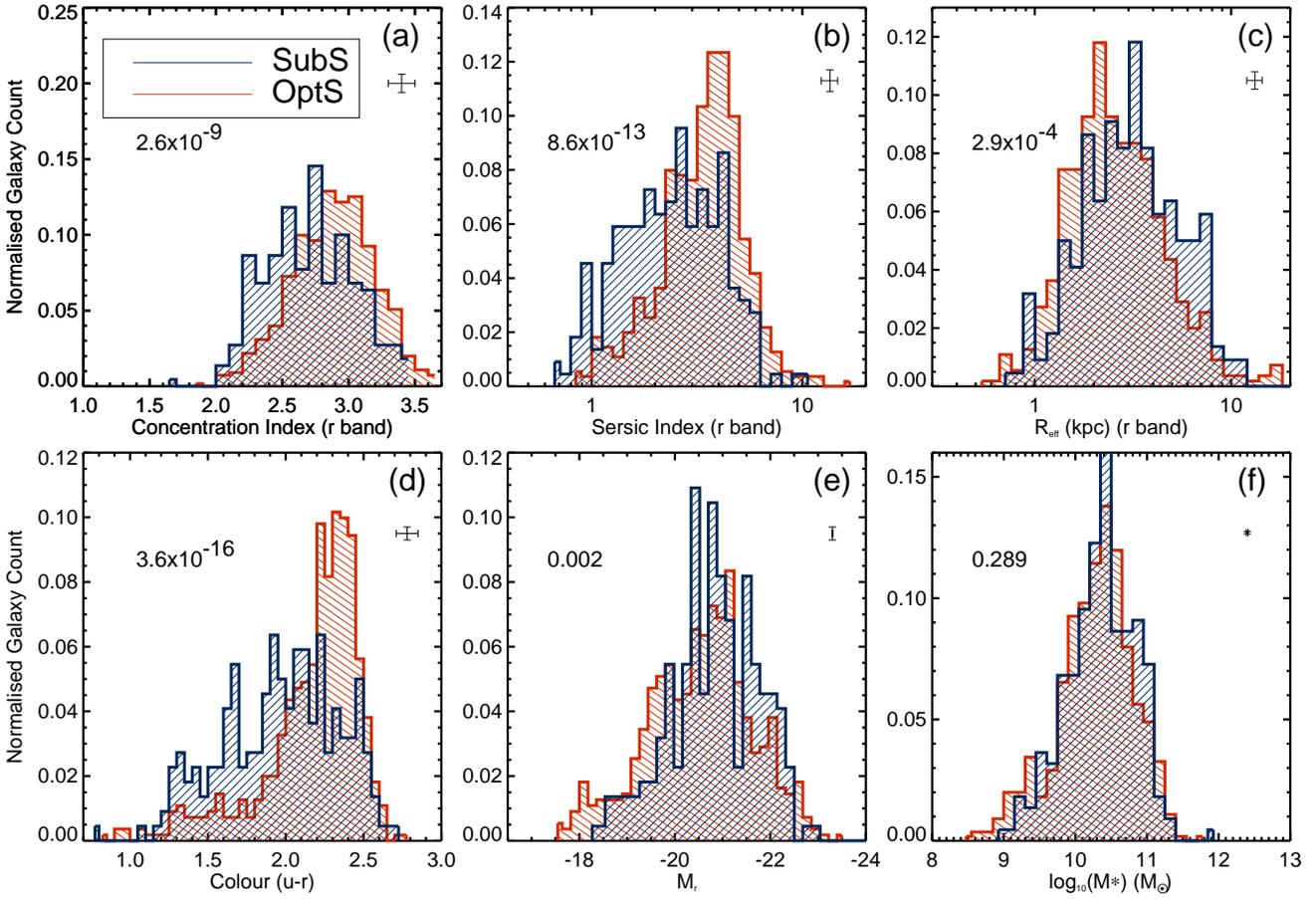} 
 \caption{Histograms representing distributions of ETGs in the \textit{SubS} and \textit{OptS} for the optical parameters discussed in $\S$ \ref{sec:sec3}. From left to right, top to bottom, we show the distributions of concentration index, S\'ersic index, intrinsic effective radius, (u-r) colour, absolute r band magnitude, and stellar mass. Plots include average errors for these parameters, as well as error bars for the normalised binning. KS-probabilities from Table \ref{tab:KS2} are also included for each set of distributions.} \label{fig:sfprops}
\end{center}
\end{figure*}	

\citet{kormendy_2009} split elliptical galaxies into two main classes: Giant Ellipticals and Normal/Dwarf Ellipticals. The primary class can be defined as having a brightness $\textrm{M}_{r}\le -21.5$ and n$\ge$4, whilst the Normal/Dwarf class has the reverse criteria. To gain understanding of the characteristics of galaxies seen in our samples, we plot distributions of multiple parameters in Fig. \ref{fig:sfprops} for both our Sub-mm and Optical samples.  At a glance, most of these distributions are seen to differ for the \textit{SubS} and \textit{OptS}. Based on the \citet{kormendy_2009} brightness definition, both samples contain a broad range of r band absolute magnitudes, indicating a mix of both giant and normal ETGs. 

To compare the structural and intrinsic properties of ETGs in our samples, we carry out a series of Kolmogorov-Smirnov (KS) tests for our samples and subsets thereof, the results of which are shown in Table \ref{tab:KS2}. We choose to test the samples as a whole because they are volume-limited, and therefore the testing is unlikely to be biased by, for example, redshift or mass effects. We also test two subsets: the first containing those galaxies classed as Giant ETGs (M$_{r}\le$-21.5) and the second classed as Normal/Dwarf ETGs (M$_{r}>$-21.5). The aim of this subset testing is to investigate whether these two so-called separate populations have their own unique set of properties for sub-mm detected and undetected ETGs. We find that 36$\%$ of Giant ETGs are sub-mm detected, whereas only a quarter of the Normal ETGs are detected.

Table \ref{tab:KS2} therefore shows parameters which appear to be most representative of the characteristics of the two samples of galaxies. We use a combination of Table \ref{tab:KS2} and Fig. \ref{fig:sfprops} to investigate how these distributions vary for our two samples. We choose the KS-test because of its usefulness as a non-parametric test for checking a null hypothesis for unbinned distributions that are functions of a single independent variable \citep{press1992numerical}. We choose KS-probabilities of 1$\%$  as the significance level for our distribution testing.

By examining the spectroscopic redshift values in this table, we can deduce that the typical redshift for the galaxies in both samples do not differ for our whole samples or Giant subset, although there is a difference in distributions for the Normal subset.

Concentration index, defined as the ratio of Petrosian r90 to r50 radii \citep{mateus_2006}, where r90 and r50 are the SDSS circular aperture radii within which 90$\%$ and 50$\%$ of the flux are contained \citep{bell_2003}, is the first parameter plotted in Fig. \ref{fig:sfprops}. It is linked to the concentration of light in the centre of a galaxy, and it has higher values in ETGs than in LTGs \citep{strateva_2001,conselice_2006}. Both Table \ref{tab:KS2} and Fig. \ref{fig:sfprops}(a) show that concentration index for galaxies in the two full samples differs significantly, with the probability of getting the null hypothesis of no difference approaching zero. The distribution for the $SubS$ is offset towards low concentrations for the full samples and for both low and high luminosity subsets. These differences indicate that H-ATLAS detected ETGs are likely to have less concentrated light distributions, and therefore may have more perturbed internal structure than those which are undetected, indicating some past merging or formation activity. 

\begin{figure}
\begin{center}
\vspace*{-0.6in}
 \hspace*{-0.15in}
   \includegraphics[width=0.52\textwidth]{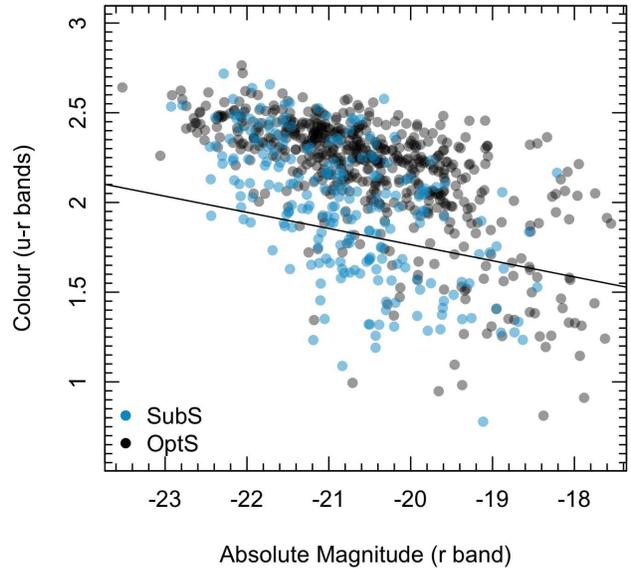} 
  \end{center}
 \caption{Scatter plot showing the colour-magnitude (CM) distribution of galaxies in the $SubS$ (blue circles) and $OptS$ (black-grey circles). The division between Red Sequence and Blue Cloud defined in the text is shown as the black solid line.}
   \label{fig:HATLASCMdiagram}
\end{figure}

	\begin{table*}
   \centering
      \begin{tabular}{l l c c c c c c c}

      \hline

&	\textbf{Parameter} &	\textbf{KS-stat (D)} &	\textbf{KS-prob}	&	\multicolumn{2}{c}{\textbf{Sample Size}}	&	\multicolumn{2}{c}{\textbf{Mean Value}}	\\
      \hline

      	& &	&	&	\textit{\textit{SubS}}	&	\textit{\textit{OptS}}	&	\textit{\textit{SubS}}	&	\textit{\textit{OptS}}\\
      \hline \hline 
	&	Redshift			&	0.122	&	0.017	&	220	&	551	&	0.044	&	0.046\\
	&	Concentration index (r band)	&	0.253	&	2.6$\times$10$^{-9}$	&	220	&	551	&	2.71	&	2.88\\
	&	S\'ersic index (r band)		&	0.298	&	8.6$\times$10$^{-13}$	& 	220	&	551	&	2.72	&	3.67\\
	&	Effective Radius (r band)	&	0.166	&	2.9$\times$10$^{-4}$	&	220	&	551	&	 4.02	&	3.13\\
(1) Full Samples	&	Colour (u-r band)&	0.336	&	3.6$\times$10$^{-16}$	&	220	&	551	&	1.97	&	2.17\\
	&	Absolute Magnitude (r band)	&	0.148	&	0.002	&	220	&	551	&	-20.84	&	-20.52	\\
	&	Stellar mass 			&	0.078	&	0.289	&	220	&	551	&	10.36	&	10.28\\
	&	UV Colour (NUV-r band)		&	0.517 	&	4.5$\times$10$^{-32}$	&	181	&	481	&	4.05	&	5.33\\
	&	Surface Density 		&	0.257	&	1.6$\times$10$^{-9}$	&	219	&	551	&	7.20	&	12.63	\\

\hline
      &		&	&	&	\textit{\textit{SubS}}	&	\textit{\textit{OptS}}	&	\textit{\textit{SubS}}	&	\textit{\textit{OptS}}\\
      \hline \hline 
	&	Redshift			&	0.116	&	0.684	&	57	&	99	&	0.047	&	0.046\\	
	&	Concentration index (r band)	&	0.362	&	7.8$\times$10$^{-5}$	&	57	&	99	&	2.93	&	3.16\\
	&	S\'ersic index (r band)		&	0.295	&	0.003	& 	57	&	99	&	3.71	&	4.41\\
	&	Effective Radius (r band)	&	0.257	&	0.013	&	57	&	99	&	 7.76	&	5.82\\
(2) Giant ETGs	&	Colour (u-r band)	&	0.363	&	9.5$\times$10$^{-5}$	&	57	&	99	&	2.28	&	2.41\\
      &	Absolute Magnitude (r band)	&	0.149	&	0.367	&	57	&	99	&	-22.01	&	-22.08	\\
	&	Stellar mass 			&	0.231	&	0.036	&	57	&	99	&	10.93	&	10.99\\
	&	UV Colour (NUV-r band)		&	0.667	&	1.5$\times$10$^{-12}$	&	50	&	88	&	4.99	&	5.33\\
	&	Surface Density 		&	0.316	&	0.001	&	56	&	99	&	4.12	&	18.22	\\
\hline	
      &		&	&	&	\textit{\textit{SubS}}	&	\textit{\textit{OptS}}	&	\textit{\textit{SubS}}	&	\textit{\textit{OptS}}\\
      \hline \hline 
	&	Redshift			&	0.154	&	0.006	&	163	&	452	&	0.043	&	0.046\\	
	&	Concentration index (r band)	&	0.292	&	1.8$\times$10$^{-9}$	&	163	&	452	&	2.64	&	2.82\\
	&	S\'ersic index (r band)		&	0.388	&	1.9$\times$10$^{-16}$	& 	163	&	452	&	2.37	&	3.50\\
	&	Effective Radius (r band)	&	0.150	&	0.008	&	163	&	452	&	 2.71	&	2.54\\
(3) Normal ETGs	&	Colour (u-r band)	&	0.413	&	1.2$\times$10$^{-18}$	&	163	&	452	&	1.85	&	2.12\\
	&	Absolute Magnitude (r band)	&	0.154	&	0.006	&	163	&	452	&	-20.43	&	-20.18	\\
	&	Stellar mass 			&	0.067	&	0.649	&	163	&	452	&	10.16	&	10.12\\
	&	UV Colour (NUV-r band)		&	0.570	&	4.6$\times$10$^{-30}$	&	138	&	396	&	3.74	&	5.33\\
	&	Surface Density 		&	0.262	&	1.0$\times$10$^{-7}$	&	163	&	452	&	8.25	&	11.52	\\
\hline
   \end{tabular}
   \caption{Two-tailed Kolmogorov-Smirnov (KS) test results for the distribution of host galaxy parameters for our ETGs, tested as follows: (1) testing the full samples, (2) testing those ETGs in the samples with M$_{r}\le$-21.5 and (3) those ETGs with M$_{r}>$-21.5. The first column gives the parameter which has been tested, with the two distributions being the sub-mm detected and undetected samples respectively. The second column gives the KS-statistic D, which is defined as the maximum value of the absolute difference between two cumulative distribution functions. The third column gives KS-probability, which is the probability for the null hypothesis that these data sets are drawn from the same distribution. The fourth and fifth columns show the numbers of \textit{SubS} and \textit{OptS} galaxies tested. The final two columns show the mean values for the \textit{SubS} and \textit{OptS} parameter distributions. The test is carried out using the IDL routine KSTWO.}
   \label{tab:KS2}
\end{table*}

We next look at S\'ersic profile properties for the samples. \citet{kelvin_2011} describe how their SIGMA wrapper around the GALFIT3 profile fitting program \citep{peng_2010} is used to fit the entire GAMA database with 2D S\'ersic profiles in multiple wavebands. The key parameters that we are interested in from this profile fitting are the S\'ersic indices and effective radii of galaxies in our samples. Initially we examine S\'ersic index, which gives us information on the distribution of light within a galaxy. This parameter is more useful than concentration index as it is less affected by seeing, having been convolved with the point-spread function (PSF), and therefore does not vary so much with redshift. The recovered S\'ersic index distributions show similar trends to those observed for concentration index. The distribution of the samples in equal log spacing shown in Fig. \ref{fig:sfprops}(b) indicates that ETGs in the $SubS$ have lower S\'ersic indices, and KS-testing the samples shows them to be significantly different. One possible interpretation of this result is that the difference found in the two samples is due to the effect of dust, which has been shown (eg. \citealp{pastrav_2013}) to lower values of measured S\'ersic index. Alternatively, these results for both S\'ersic and concentration index may be caused by morphological disturbances, or problems with the morphological selection process, which is limited by the resolution of the SDSS.

It is interesting to note that mean S\'ersic indices for the two samples are higher for the Giant ETGs subset than the Normal ETGs subset, as is predicted by the \citet{kormendy_2009} separation criteria. In addition, the observed differences in distributions of S\'ersic index affects mainly the Normal ETG subset (see Table \ref{tab:KS2}). Thus, if we believe that this difference is caused by dust, S\'ersic index would appear to be most affected by its presence in low luminosity ETGs. This could indicate that higher luminosity ETGs contain lower normalised dust masses, leading to higher S\'ersic indices and profiles more similar to $OptS$ ETGs. We explore this effect with respect to specific dust mass in $\S$\ref{sec:sec4}.

Apparent effective radii are also calculated for the galaxies through this S\'ersic profile fitting, and we convert these to intrinsic radii using simple geometry and angular diameter distances. The distributions are plotted in log space in Fig. \ref{fig:sfprops}(c). It appears that $OptS$ ETGs typically have smaller effective radii than those in the $SubS$, although this result is not robust when the sample is divided into luminosity subets.

We examine (u-r) colours of both samples, where galaxies in the optical Red Sequence (RS) are separated from the Blue Cloud (BC) using the following relation:

\begin{equation}
\textrm{u-r} = -0.09M_{r}-0.0347
\end{equation}

\noindent This was calculated by fitting approximate straight lines to the GAMA RS and BC, and finding the line equidistant in magnitude from these fits. This is a similar method to other published works for different optical colours (e.g. \citealp{bell_2003}). ETG colours are explored in Fig. \ref{fig:HATLASCMdiagram}, where the (u-r) CM scatter plot is shown for both samples. The diagram highlights the point that ETGs, whether detected in the sub-mm or not, have a multitude of both blue and red colours. Interestingly, the $SubS$ does not show as dense a clustering in the Red Sequence as the $OptS$ and is more evenly distributed towards the blue end. Both Figs. \ref{fig:sfprops}(d) and \ref{fig:HATLASCMdiagram} indicate that the $SubS$ contains a larger proportion of blue ETGs. Table \ref{tab:KS2} also shows the distribution of optical colours is quite different, even when testing separate subsets.

The full sample testing of absolute magnitudes indicates different distributions. However, results from subset testing indicate that the Giant ETGs have similar distributions, whilst Normal ETGs are different with $\sim$0.6$\%$ probability. This is unsurprising, given the $OptS$ ETGs extend to fainter magnitudes than $SubS$ ETGs, shown in Fig. \ref{fig:HATLASCMdiagram}.

Given the link between luminosity and stellar mass of galaxies, we would expect equivalent results for stellar mass distributions. However, all the samples have similar mass  distributions, although once again the result is less robust for the Giant subset. On average, the $SubS$ reaches to brighter absolute magnitudes than the $OptS$, whereas mass differences between the samples are not significant. This may mean a decrease in mass-to-light ratios for $SubS$ ETGs.

Altogether, after testing these intrinsic properties, we find sub-mm detected ETGs to be bluer than undetected ETGs (see also results by \citealp{dariush_2011}), and we also show they are likely to have similar masses, lower S\'ersic indices, are less centrally concentrated and less compact than undetected ETGs in our samples.

\subsection{UV Parameters}\label{sec:UV}

Availability of GALEX data means that we can look at the ultraviolet (UV) bands for both our \textit{OptS} and \textit{SubS}, typically used as an indicator of recent star formation. The GALEX bands are Far UV (FUV), with an effective wavelength of 1528$\textrm{\AA}$, and Near UV (NUV), with an effective wavelength of 2271$\textrm{\AA}$ \citep{martin_2005}. These GAMA-GALEX catalogues seek to reconstruct the original UV flux for each given optical source, and include UV fluxes and AB apparent magnitudes for both wavebands.%Have another look at this paragraph when have a chance.

We corrected NUV CATAID-matched apparent magnitudes for Galactic extinction (for more details, see \citealp{wyder_2007}) and converted them to absolute magnitudes using GAMA database k-corrections derived from \textit{kcorrect v.4.2} software, described in \citet{blanton_2007}. This IDL code calculates K-corrections for the galaxies based on the best fit sum of templates to an SED and accounts for both the angular diameter distance and cosmological surface-brightness dimming for the distance modulus.

\begin{figure}
\begin{center}
\hspace*{-0.22in}
 \includegraphics[width=0.53\textwidth]{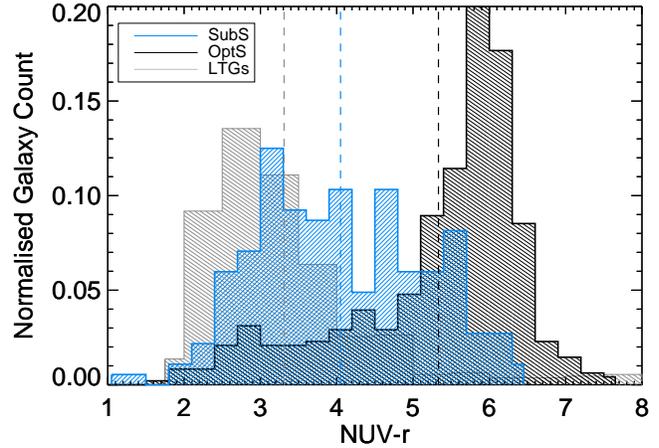} 
 \end{center}
 \caption{Histograms of UV-Optical colour for \textit{OptS} (black) and \textit{SubS} (blue), as well as H-ATLAS detected LTGs (grey). The \textit{SubS} appears to be skewed towards the bluer end, whilst the peak of the \textit{OptS} is decidedly within the red end of the plot. The blue, black and grey dot-dashed lines show the respective mean colours for the \textit{SubS}, \textit{OptS} and LTGs (4.05, 5.33 and 3.31 mags).}
   \label{fig:NUVCM}
\end{figure}

We obtain 481 and 184 galaxies with GALEX NUV detections in the \textit{OptS} (87$\%$) and \textit{SubS} (83$\%$) respectively. The $SubS$ is brighter on average in the NUV by $\sim$2 mags. We examine the NUV-r colours in Fig. \ref{fig:NUVCM}, where we see the \textit{SubS} contains bluer UV-Optical colours than the \textit{OptS}, similar to the H-ATLAS result for ETGs in the SDP field of \citet{Rowlands_2011}. As in $\S$\ref{sec:sec3}, we perform KS-tests on the full $SubS$ and $OptS$ samples and their luminosity subsets. The results of these tests are presented in Table \ref{tab:KS2} and show the distributions of UV colours are very different for ETGs in all these cases, with $SubS$ ETGs showing much bluer colours overall.

For comparison, we also show where H-ATLAS detected LTGs (visually classified in the same process as ETGs in our samples) lie on this histogram. The colours of these LTGs are bluer on average than both our ETG samples, but the $SubS$ straddles the gap between blue and red modes filled by LTGs and $OptS$ respectively. This indicates that H-ATLAS detected ETGs are forming a colour population of their own.

\begin{figure}
\begin{center}
\hspace*{-0.3in}
 \includegraphics[width=0.53\textwidth]{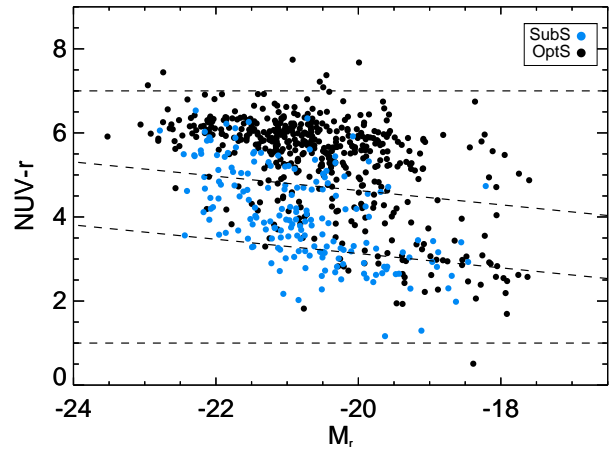} 
 \end{center}
 \caption{The distribution of $SubS$ (blue filled circles) and $OptS$ (black filled circles) on the UV-optical colour-magnitude diagram. Dashed lines indicate the cutoffs taken from \citet{bourne_2012} to separate galaxies into Red Sequence, Green Valley and Blue Cloud respectively.}
   \label{fig:NUVdiag}
\end{figure}

We next examine the UV-optical CM diagram in Fig. \ref{fig:NUVdiag}, and follow the method of \citet{bourne_2012}, using NUV-r colour boundary cuts shown in Eqs. \ref{eq:colourcut1}, \ref{eq:colourcut2} and \ref{eq:colourcut3} below to define regions as the Red Sequence (NUV-RS), Green Valley (NUV-GV) and Blue Cloud (NUV-BC) respectively.

	\begin{table}
   \centering
      \begin{tabular}{l c c c c }
      \hline
      \textbf{Colour Region} &	\multicolumn{2}{c}{\textbf{\textit{SubS}}} &	\multicolumn{2}{c}{\textbf{\textit{OptS}}}	\\
      \hline
      	&		\textit{Galaxies}	&	\textit{Fraction}	&	\textit{Galaxies}	&	\textit{Fraction}\\
      \hline \hline 
	Red Sequence	&	50	&	0.27	&	373	&	0.78\\
	Green Valley	&	88	&	0.48	&	77	&	0.16\\
	Blue Cloud	&	46	&	0.25	&	31	&	0.06\\
\hline

   \end{tabular}
   \caption{Numbers and Fractions of ETGs residing in the UV-Optical Red Sequence, Green Valley and Blue Cloud, as defined by \citet{bourne_2012}.}
   \label{tab:Colours}
\end{table}

\begin{eqnarray}
  1.23 - 0.17M_{r} &\le& (NUV-r)_{rest} \le 7.0  \label{eq:colourcut1}\\
 -0.27 - 0.17M_{r} &\le& (NUV-r)_{rest} < 1.23 - 0.17M_{r} \label{eq:colourcut2}\\
  1.0 &\le& (NUV-r)_{rest} < -0.27 - 0.17M_{r}	\label{eq:colourcut3}
\end{eqnarray}

\noindent The results of applying these boundary conditions to define different regions are shown in Table \ref{tab:Colours}. This highlights a key difference between the \textit{SubS} and \textit{OptS}. There is a large fractional difference between the populations in all these regimes, with $OptS$ galaxies most prominently located in the NUV-RS, and $SubS$ ETGs mostly dominated by the NUV-GV. It is difficult to determine whether the 48$\%$ sub-mm detected galaxies in the NUV-GV are ETGs in transition from the Blue Cloud to the Red Sequence, or whether they are NUV-BC ETGs that are dust-reddened to appear in the NUV-GV, although absolute NUV flux levels do suggest ongoing or recent star-formation in $SubS$ ETGs. 

The sub-mm detected ETG occupation of the NUV-BC is a factor of three more than that of undetected ETGs, possibly indicating more ongoing or recent star formation in the former. If the UV emission in ETGs is dominated by emission from recent star formation and dust is predominantly heated by photons from the young stellar population, then we would expect $SubS$ ETGs to be dust-reddened and driven towards the NUV-RS. The fact that the $SubS$ ETGs occupy mostly the NUV-GV implies that one of these assumptions may not hold. Additionally, we might predict that the dust does not lie within the same geometry as the star formation, potentially implying star formation is not the main source of heating the dust.

\subsection{Environment Parameters}

The \citet{dressler_1980} morphology-density relation is a key driver for investigating the varying environments of ETGs. Although it has been shown that bulge-dominated galaxies are more commonly found in the densest regions of the Universe, it is unclear whether ETGs with differing dust properties are typically found within different environments.

Disk-dominated galaxies are thought to be transformed into bulge-dominated galaxies via processes which occur in dense environments (see \citet{boselli_2006} for a review of such processes). This work is only concerned with bulge-dominated galaxies, or ETGs. However, we are comparing sub-mm non-detected passive ETGs, which may likely be at the end stages of their lives (\textit{OptS}) with dusty ETGs which are likely to have active star-formation (\textit{SubS}). It would be interesting to see whether these two classes of ETGs reside in different environments.

The surface density information used to investigate environments for our ETG samples was taken from the GAMA EnvironmentMeasures database. These surface densities ($\Sigma_{gal}$) in galaxies Mpc$^{-2}$ were calculated based on the Nth Nearest Neighbour method, using:

\begin{equation}
\Sigma_{gal} = \frac{N}{\pi D_{N}^{2}}
\end{equation}
	
\noindent where N is the nearest neighbour number, set at a value of 5, and D$_{N}$ is the projected distance in co-moving Mpc to the N$^{th}$ nearest neighbour within a velocity cylinder $\pm$1000kms$^{-1}$ from a volume-limited, density-defining population. This population is defined by an r band absolute magnitude M$_{r}\le$-20 and a redshift selection of 0.002$\le$z$\le$0.18 \citep{wijesinghe_2012,brough_2012}. We choose to include upper limits calculated based on the distance to the nearest angular edge. Only one galaxy within our two samples is removed because the density cannot be estimated, reducing the $SubS$ to 219 ETGs. We find surface density to be an extremely useful parameter for us to work with because it gives us information about the area around the galaxy and whether there are likely to be any real interactions occurring between it and its neighbours.

	\begin{figure}
\begin{center}
\hspace*{-0.2in}
 \includegraphics[width=0.52\textwidth]{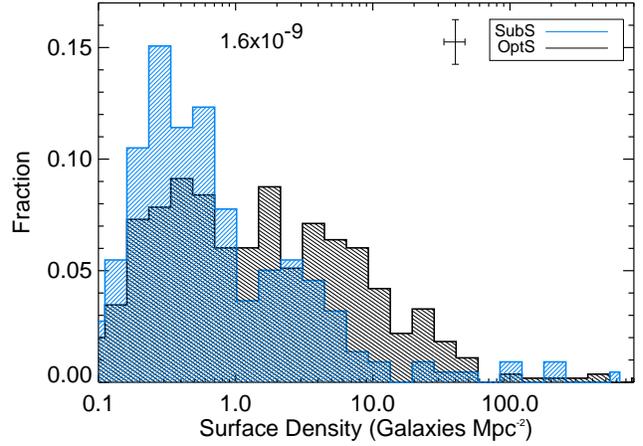} 
 \caption{Normalised histograms of surface density in log bins, where the blue histogram is the $SubS$ and black histogram is the $OptS$. The error bars represent average errors on both surface density and the normalised binning. KS-probability for comparison of surface density distributions from Table \ref{tab:KS2} is also included.} \label{fig:density}
\end{center}
\end{figure}	

Fig. \ref{fig:density} shows the surface densities for both the reduced \textit{SubS} and \textit{OptS}. The \textit{OptS} is more distributed over the range of surface densities, indicating more of these galaxies are in denser regions. As in $\S$\ref{sec:sec3}, we KS-test the distributions of densities for the samples and within luminosity subsets. Results given in Table \ref{tab:KS2} show that the ETGs have different density distributions in the main sample, as well as both luminosity subsets, with probabilities of below 1$\%$.

\citet{robotham_2011} produced a GAMA Galaxy Group Catalogue (G$^{3}$C) that identifies which groups the GAMA galaxies belong to, as well as assigning each group a set of properties. It is assumed that if the galaxies are not assigned a Group ID, then they are in the field. We have used G$^{3}$C to assign our galaxies either field or group status by splitting the \textit{OptS} and \textit{SubS} into two: grouped and ungrouped galaxies. We then compare the two samples and find that the \textit{SubS} appears to have galaxies which are almost equally split between the field (52$\%$) and groups (48$\%$), whereas the $OptS$ shows more of a difference with 62$\%$ in the field and 38$\%$ in groups. ETGs within the groups can be further split according to the multiplicity of the group: ETGs in small groups with less than five galaxies (32$\%$ and 24$\%$ for the $SubS$ and $OptS$, respectively), and larger groups with five or more galaxies (15$\%$ and 14$\%$ respectively).

\begin{figure}
\begin{center}
\hspace*{-0.25in}
 \includegraphics[width=0.52\textwidth]{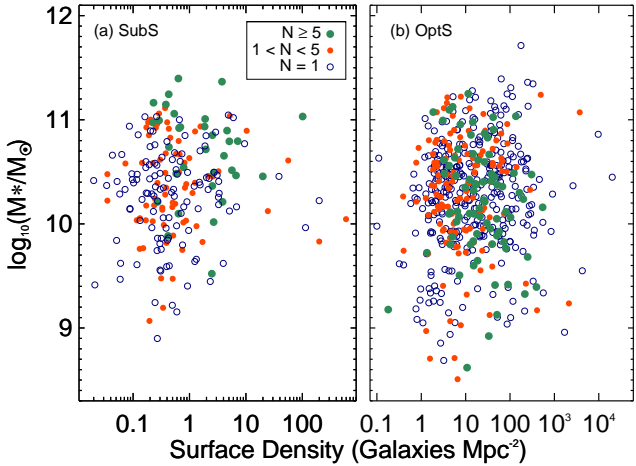} 
 \caption{Variation of stellar mass with surface density. (a) shows results for $SubS$ and (b) for $OptS$. Both plots are coloured by group multiplicity: open blue circles are galaxies which are ungrouped and assigned a group size of 1, red filled circles represent ETGs in small groups ($<$ 5 galaxies) and large, green filled circles those in large groups ($\ge$ 5 galaxies). Note that the horizontal scales differ in these two panels.} \label{fig:densitymass}
\end{center}
\end{figure}

Fig. \ref{fig:densitymass} shows the distribution of surface densities with stellar mass for our two samples. The galaxies have been coloured by the multiplicities of their groups, making it simple to identify lone ETGs, ETGs in small groups and those in large groups. Both plots show that higher densities are linked with larger groups, but are not exclusively dominated by these larger groups. Both of these plots also show there are some ETGs with high surface densities which do not seem to share a group. This may be an effect of the groups data selection function, which changes with redshift. This figure highlights the similar $SubS$ and $OptS$ mass ranges, but different galaxy surface density ranges. Testing for Spearman correlations does not reveal any trends between surface density, group multiplicity and stellar mass, although these same tests suggest that the $OptS$ contains more ungrouped ETGs in high density areas (r$_{s}$=0.005) than the $SubS$ (r$_{s}$=0.24).

There also appear to be some missing galaxies in Fig. \ref{fig:densitymass}(a). The $OptS$ shows galaxies of quite low stellar mass (M$_{*}<$10$^{10}$M$_{\odot}$) going to higher densities, but these do not appear for the $SubS$. In general, H-ATLAS is known to preferentially detect higher mass galaxies, which could partially explain this. In addition, the larger H-ATLAS PSF when compared to that of the optical may result in low counterpart reliability for ETGs in the densest regions, resulting in their removal from our samples. The sparsity of low mass, dusty ETGs at high surface densities may be caused by sample completeness effects. However, the sparsity of dusty ETGs in general at high densities suggests that dusty ETGs do not occupy the densest regions of groups and clusters, unlike their non-dusty counterparts. \citet{kauffmann_2004} showed that generally, massive galaxies in low density environments in their SDSS sample contain the most dust, based on evidence of optical attenuation.  \citet{kaviraj_2011} studied optical images of ETGs with visible dust lanes and patches. They found that dusty ETGs occupy less dense environments than those with no sign of dust obscuration. Our findings are qualitatively consistent with these results.

\section{Dust Properties of Detected ETGs}\label{sec:sec4}

\subsection{Fitting Modified Planck Functions}\label{sec:dustydust}

	\begin{figure*}
\begin{center}
 \includegraphics[width=\textwidth]{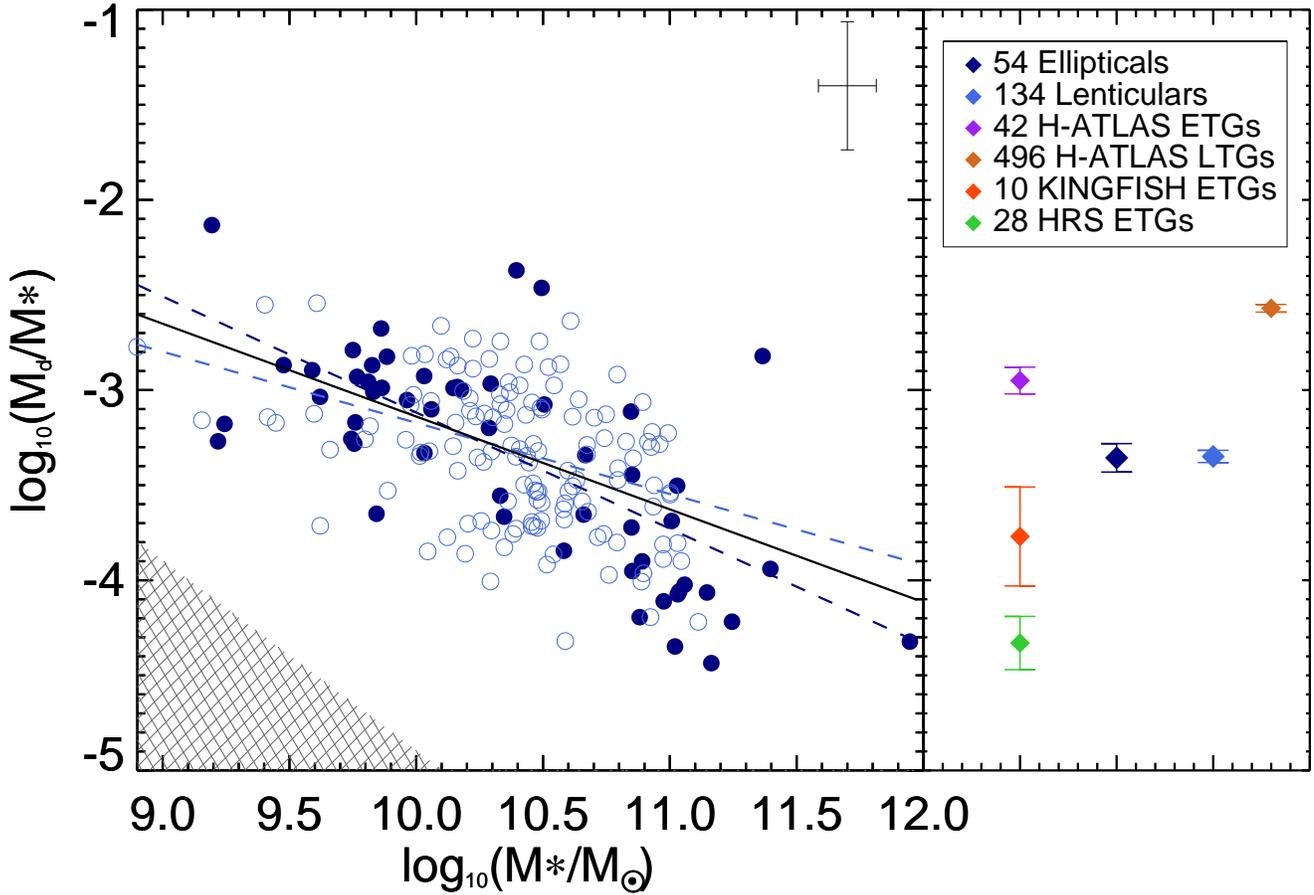} 
 \caption{Specific dust mass of ETGs in our sample $SubS$ as a function of stellar mass. The points are coloured by morphology: dark blue filled circles for ellipticals (E) and light blue open circles for lenticulars (S0a). Average errors are shown as bars on the top right. Same colour dashed lines represent best fit straight lines through the respective points. The black solid line is the overall fit to all points. The 250$\mu$m flux limit for a typical galaxy with a upper limit dust temperature of 30K is indicated by the grey cross-hatched region. In the panel to the right, filled diamonds and error bars represent mean specific dust masses for a range of studies with Herschel data, described in the text.} \label{fig:dustmass}
\end{center}
\end{figure*}

	The $SubS$ ETGs have both $Herschel$ PACS 100 and 160$\mu$m measurements, as well as 250, 350 and 500$\mu$m fluxes from the SPIRE instrument. We select a subsample (3$\sigma$ sample) of 188 ETGs which have at least 3$\sigma$ flux levels (22.6 mJy) in the 350$\mu$m SPIRE waveband, as well as the requisite 5$\sigma$ emission in at least one SPIRE sub-mm waveband (see $\S$\ref{sec:SelEff}). PACS measurements are missing for 27 of these galaxies, and for these we choose to fit only SPIRE flux densities. As specified for all H-ATLAS Phase 1 data, flux errors of 7$\%$ of the catalogue flux values are added in quadrature to account for the uncertainty in the SPIRE photometric calibration, and 10$\%$ to the PACS fluxes.
	
	In order to estimate dust temperature and mass we then fit isothermal, modified Planck functions to the data for each galaxy, of the form:
	
	\begin{equation}
	F_{\lambda} = \Omega B_{\lambda}\lambda^{-\beta}
	\end{equation}
	
	\noindent where $B_{\lambda}$ is the Planck function, $\Omega$ is the amplitude parameter for the model fit, and $\beta$ the dust emissivity index which we fixed as 2.0 (e.g. \citealp{galametz_2011,davies_2012,cortese_2012}). This greybody model is fit to the FIR/sub-mm flux densities, which are corrected for redshift. We are thereby able to measure the typical rest-frame temperatures of emission by the dust distributions in these galaxies. 
	
	Dust masses for this subsample are calculated using
	
	\begin{equation}
	M_{d} = \frac{F_{250}D_{L}^{2}K}{\kappa_{250}B(T)_{250}\left(1+z\right)}\label{eq:dustmass}
	\end{equation}
	
	\noindent (\citealp{whittet_1992}, Eq. 6.12), where F$_{250}$ is the observed flux value in the 250$\mu$m waveband in Janskies (Jy), D$_{L}$ the luminosity distance to the source and K is the K-correction as defined in Eq. 2 of \citet{dunne_2010}. The assumed mass absorption coefficient ($\kappa_{250}$) is 0.89m$^{2}$kg$^{-1}$ at 250$\mu$m \citep{dunne_2010}. The value of B(T)$_{250}$ in Eq. \ref{eq:dustmass} is calculated using the best fit, cold dust, rest-frame temperature from the SED fitting for each galaxy. We choose to measure the dust mass using this waveband because of the relatively smaller errors when compared with the other SPIRE wavebands.

The results of our fitting give the mean cold dust temperature for the ETGs as 22.1$^{+2.7}_{-2.0}$K with 5$^{th}$ to 95$^{th}$ percentiles of 14-28K, with no obvious variation of fit temperature with stellar mass. The mean cold dust mass is calculated as (1.8$\pm$0.5)$\times$10$^{7}$M$_{\odot}$ with a percentile range of (0.19-5.41)$\times$10$^{7}$M$_{\odot}$. The resultant average specific dust mass (log$_{10}$(M$_{d}$/M$_{*}$)) is calculated as -3.37. Errors on temperature are given as formal uncertainties and propagated through Eq. \ref{eq:dustmass}. Due to the non-linearity of this propagation, we only show average dust mass errors for those galaxies with full PACS and SPIRE data. Thus the correlation between dust temperature and mass is accounted for within dust mass errors.

We present our specific dust masses in Fig. \ref{fig:dustmass}, plotted against stellar mass. Overall, the galaxies show a clear trend for lower mass galaxies to contain higher normalised dust masses, with a straight line fit to the distribution yielding a Pearson coefficient (r$_{p}$) of -0.55. The dependancy on morphology within ETGs is addressed in $\S$\ref{sec:ellipticals}. Here we compare our average normalised dust mass with those from other works with $Herschel$. These results and the standard errors on the mean are also shown in Fig. \ref{fig:dustmass}(right) for ellipticals and lenticulars separately.

Our ETG sample does not contain many low mass ETGs, which indicates that there may be missing galaxies at the low stellar mass and specific dust mass end of Fig. \ref{fig:dustmass}. To test for this, we show a cross-hatched region on the plot, representing the 250$\mu$m 5$\sigma$ limit. This region is based on the same dust mass calculations as our ETGs, with a representative upper limit dust temperature of 30 K and minimum redshift of 0.013. This excluded region becomes larger with decreasing temperature and increasing redshift, and represents the region of the plot which H-ATLAS is insensitive to. This indicates that we are unlikely to be missing ETGs on this diagram due to the flux limit, since we would expect to detect an increasing proportion of galaxies above this limit.

Based on the results found in $\S$\ref{sec:sec3}, we want to further explore the properties of the ETGs for high and low luminosity subsets. Luminosity and stellar mass are known to be linked, and we plot these properties to find that values of M$_{r}$=-21.5 mag / M$_{*}$=10$^{10.2}M_{\odot}$ can be used as a high and low luminosity/stellar mass divider. From Fig. \ref{fig:dustmass} we see that while lenticulars dominate the median range of stellar masses, ellipticals appear to be separated into low/high stellar masses. Those ETGs with the highest stellar masses show the lowest specific dust masses. We have shown that the brightest ETGs contain S\'ersic indices most similar to undetected ETGs (see $\S$\ref{sec:sec3}) and this therefore indicates that ETGs with low specific dust masses have high S\'ersic indices and are not dissimilar to ETGs in the $OptS$ in this respect.

	\begin{figure}
\begin{center}
\hspace*{-0.17in}
 \includegraphics[width=0.58\textwidth]{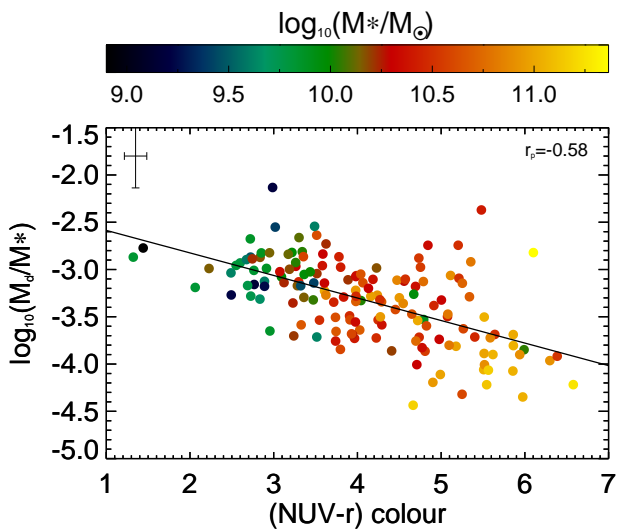} 
 \caption{Distribution of specific dust masses with (NUV-r) colour for the $SubS$. Points are coloured by the log of their stellar mass. The solid black line is the best fit straight line through the points, with the Pearson coefficient for the fit labelled in the top right-hand corner. Typical error bars for the points are shown in the top left corner.} \label{fig:NUVdust}
\end{center}
\end{figure}

We next compare results with other H-ATLAS data. Recent multi-wavelength SED fitting by \citet{Rowlands_2011} on a sample of 42 detected ETGs in the H-ATLAS SDP gave average cold dust masses as 5.5$\times$10$^{7}$M$_{\odot}$ for an allowed distribution of temperatures between 15-25K. Improving upon \citet{Rowlands_2011}, we have chosen to fit the ETGs with 3$\sigma$ detections at 350$\mu$m, and we have chosen a more straightforward single-component fitting approach. This low-redshift $SubS$ alone is five times the number of galaxies in the SDP and therefore a more statistically significant result. Our results indicate lower mean specific dust masses for ETGs than in \citet{Rowlands_2011}, which may be partly due to the nearer redshift limit of our $SubS$ sample presented here. For the sake of comparison, we have also shown the average normalised dust mass for their larger sample of LTGs. This is clearly a lot higher than the ETGs, as is to be expected for spiral galaxies, which are known to have high dust masses.

We can also compare our results with those from the KINGFISH survey, where \citet{skibba_2011} also calculate dust temperatures and masses for a sample of 10 nearby ETGs using single-temperature modified blackbodies. Their specific dust masses are lower than ours, but fall well within the range of masses exhibited by our galaxies. Therefore we can attribute the difference in mean masses to sample size. This is less of an issue with the \citet{smith_HRS} study in the HRS, where they select a sample of 62 ETGs, mainly in the Virgo Cluster, of which 45$\%$ are detected in the sub-mm. They also calculate dust masses for the 28 detected ETGs using single-temperature greybody fitting, and their average specific dust mass for detected ETGs is much lower than any of the aforementioned works, including ours. However, they do not seem to sample as large a range of environments as these other works, since their sample is restricted mainly to galaxies in the high densities of the Virgo Cluster. Further HRS work by \citet{cortese_2012} shows that in clusters such as Virgo, the dust fraction of galaxies of a given mass is significantly lower than that of galaxies in isolation or in small groups. This bias for dusty ETGs to reside in lower density environments is also hinted at in Table \ref{tab:KS2} and Fig. \ref{fig:densitymass} for our $SubS$. The KINGFISH and HRS studies look at galaxies in the very nearby universe, and therefore they are able to detect the lowest dust masses in galaxies. Additionally, \citet{smith_HRS} emphasise that they only study the most massive ellipticals in the nearby universe. Given this information, the differences in average normalised dust masses between this and these other two $Herschel$ works appear quite small.

Next we consider how the computed dust masses vary with UV-optical colours. This is presented in Fig. \ref{fig:NUVdust}, where the galaxies are also coloured by stellar masses. Here we identify a very clear trend. As our galaxies get bluer, they also increase their specific dust masses and decrease in stellar mass. This highlights a key result: the most massive ETGs are redder and therefore more quiescent, and they also contain proportionally less dust. This strengthens the link between dust and star formation and, assuming the more quiescent ETGs do not host many young blue stars, predicts that low mass red stars do not contribute as highly towards the dust presence in ETGs as bright young stars. This sort of evolution has been shown before: \citet{bourne_2012} show a similar strong trend for their sample of red galaxies. To further understand it, we would have to work out whether we are looking at these ETGs at different ages, and therefore different stages of their evolution, or whether we are looking at two completely different populations and types of ETGs.

\subsection{Contamination Issues}\label{sec:contamination}

\citet{rigby_2011} found that a significant number of sources may have 350 and 500$\mu$m flux densities that are overestimated by a factor of $\sim$2 in the H-ATLAS SDP fields. This is due mainly to source confusion where the signal-to-noise levels are low. This would result in SED fitting underestimating dust temperatures and overestimating dust masses. There is nothing that can be done to remedy this on an object-by-object basis, but we should be aware of such an effect having carried through to the Phase 1 data.

We consider whether any of our ETGs are radio-emitting AGN by referring to the Very Large Array FIRST survey, which cover 10,000 degrees over the North and South Galactic Caps, currently with 30$\%$ optical counterparts in the SDSS \citep{becker_1994} and covering the GAMA equatorial regions down to the SDSS DR6 brightness limit. We searched for potential counterparts within a 20$^{\prime\prime}$ search radius and found two of our $SubS$ ETGs, within search radii of 18.9$^{\prime\prime}$ and 9.4$^{\prime\prime}$, and 1.4 GHz integrated flux densities of 1.17 and 3.59 mJy respectively. Given the likelihood of these being false counterparts and the low associated flux densities, we deem radio emitting AGN to not be an issue for our ETGs.

If the ETGs do host radio-emitting AGN which we have been unable to detect through FIRST counterpart matching, it is necessary to consider their implications in our results. Synchrotron emission from the AGN would be modelled as a radio power-law, which may extend into the longest sub-mm wavebands. Our greybody fit parameters will then represent not only the thermal FIR/sub-mm emission, but also the synchrotron emission component. This could lead to false cooler temperatures resulting from our fits, as the apparent greybody peak would be pushed to longer wavelengths.

 We double-check whether radio-loud AGN are present in our ETGs by looking at the SPIRE 250/350 and 350/500 colours. \citet{boselli_2010} show that SPIRE colours are useful for discriminating thermal from synchrotron emission in radio galaxies, with the colours becoming very small for those galaxies with synchrotron emission. The low redshift sample we are fitting in this section have high (f350/f500$>$1) SPIRE ratios, providing further evidence that the 500$\mu$m waveband is unaffected by synchrotron emission in our sample.
 
 Free-free emission (thermal bremsstrahlung) from ionised HII regions is also a potential contamination factor in the 500$\mu$m waveband. However, because ETGs are mostly quiescent, such emission from these regions is likely to be small and is difficult to quantify.

As discussed in $\S$\ref{sec:SelEff}, we used the \citet{negrello_2010} and \citet{Gonzalez-Nuevo:2012ys} selection criteria to remove nine galaxies which are quite likely to be lensed. These galaxies did have high enough flux emission to run greybody fits across them. The results of these fits were a range of dust masses between 10$^{7-8}$M$_{\odot}$ (-3.1$<$log$_{10}$(M$_{d}$/M$_{*})<$-1.9) and temperatures between 20-35K, with only one lens galaxy candidate given a temperature as low as 11K. Although these galaxies are not included within our dust mass plots, upon inspection we note that if they were included they would not change the straight line fits in Figs. \ref{fig:dustmass} and \ref{fig:NUVdust} or affect the results significantly.

	\begin{figure*}
\begin{center}
 \includegraphics[width=\textwidth]{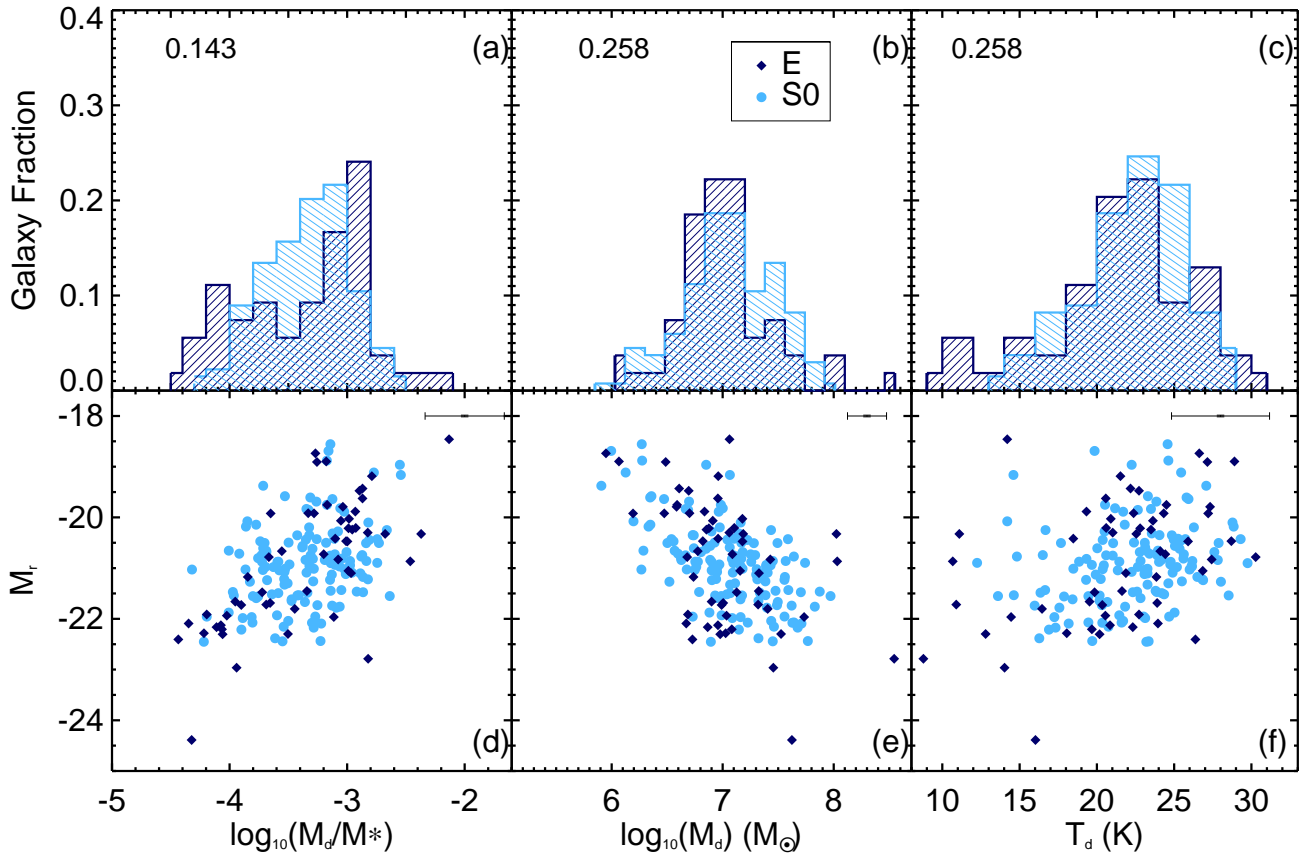} 
 \caption{Scatter plots and normalised histograms for $SubS$ ellipticals and lenticulars. Ellipticals are represented by dark blue filled circles and filled histograms, and lenticulars by light blue filled circles and histograms. From left to right: (a) distributions of dust mass normalised by stellar mass, (b) histograms of logged dust mass, (c) histograms of rest-frame dust temperatures calculated by modified Planck function fitting, (d) absolute r band magnitude (M$_{r}$) against normalised dust mass, (e) M$_{r}$ against logged dust mass, and (f) M$_{r}$ plotted against best fit dust temperature. Average errors are shown in the top-right corners. KS-probabilities described in the main text are also included for each histogram.} \label{fig:multi1}
\end{center}
\end{figure*}

	Finally, as a check, we re-ran the greybody fitting on the ETGs for just the four shortest waveband fluxes, thereby excluding the 500$\mu$m points. The resultant rest-frame temperatures did not change greatly for many of the galaxies, except in the cases ($\sim$4$\%$) where flux from the 250$\mu$m band was much greater than the 350$\mu$m flux. The lack of a strong systematic bias occurring when running this test gives further evidence that the dust characteristics of our $SubS$ are not significantly affected by synchrotron emission or other contamination issues affecting the 500$\mu$m band fluxes.

\subsection{Elliptical vs Lenticular Dust Characteristics}\label{sec:ellipticals}

	\begin{figure*}
\begin{center}
 \includegraphics[width=\textwidth]{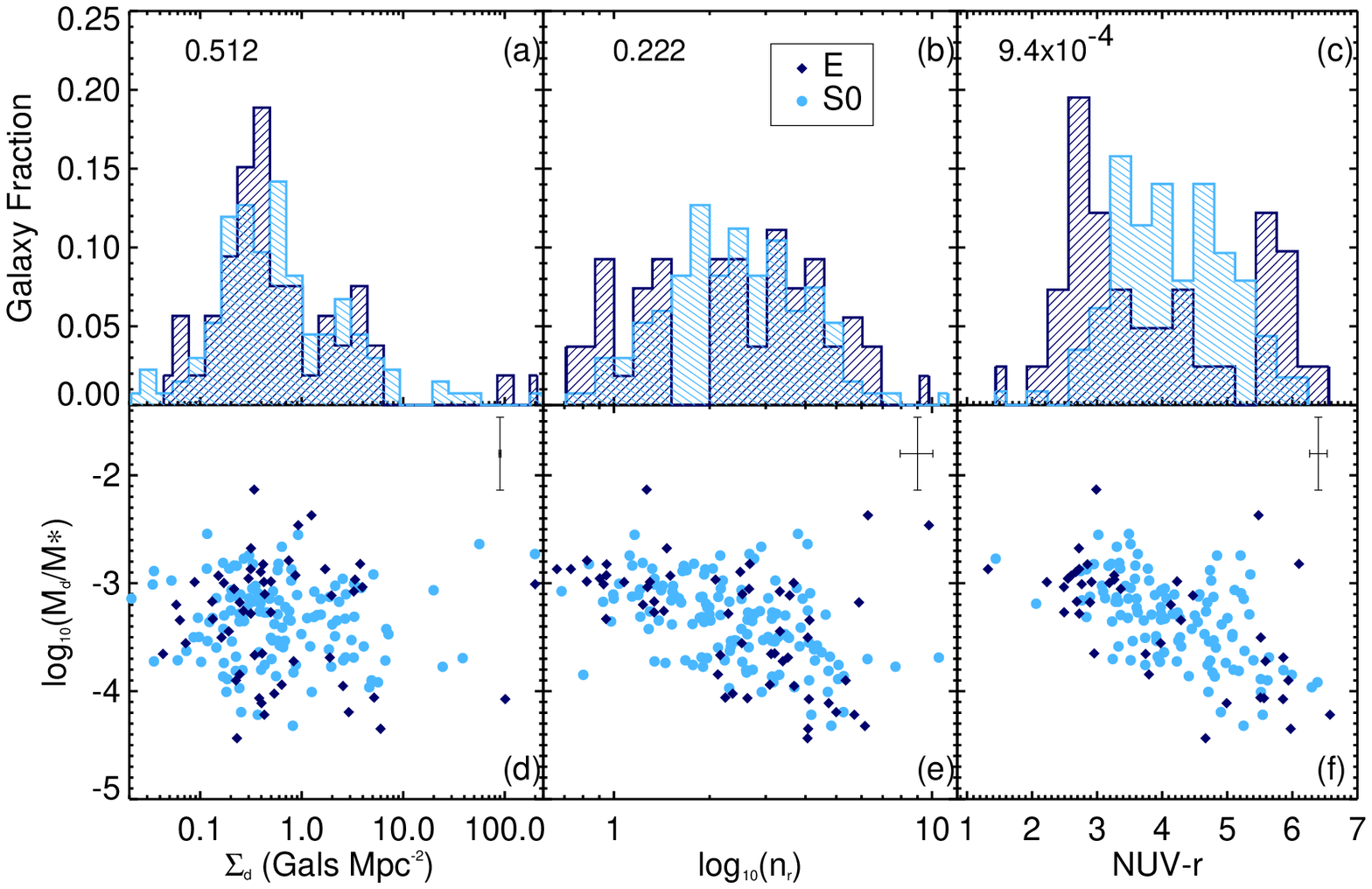} 
 \caption{Scatter plots and normalised histograms for $SubS$ ellipticals and lenticulars. Ellipticals are represented by dark blue filled circles and filled histograms, and lenticulars by light blue filled circles and histograms. From left to right: (a) normalised logged distributions of surface density $\Sigma_{d}$, (b) logged histogram of r band S\'ersic index, (c) distributions of UV-optical colour (NUV-r), (d) scatter plot showing specific dust mass against surface density, (e) specific dust mass as a function of S\'ersic index, and (f) as a function of NUV-r colour. Average errors are shown in the top-right corners. KS-probabilities described in the main text are also included for each histogram.} \label{fig:multi2}
\end{center}
\end{figure*}

It is necessary to consider the implications that S0a galaxies have created by being included within our samples, as they may have had an effect on dust results.

It is possible that lenticular galaxies are in fact Spirals which have had their star-formation cut off (e.g. \citealp{aragon_2006}). Supposing ellipticals are formed in a different way, possibly via mergers, the two different formation methods will result in different proportions of dust left within the galaxy systems. If this is the case, we are left with one of three scenarios. Either the dust masses and temperatures we are fitting to our sample are representive of dusty ETGs overall, or they are representative of either dusty S0a galaxies or E galaxies separately. 

Fig. \ref{fig:dustmass} shows the trend of normalised dust mass with stellar mass for all ETGs, coloured by morphology. We plot these separately because lenticular and elliptical galaxies are not necessarily at the same stage of their lives, and may have different evolutionary patterns. Additionally, the presence of a disk in lenticular galaxies points towards them possibly having higher dust masses than ellipticals, thus skewing our perception of the dust contents of elliptical galaxies. Such a result is not indicated in Fig. \ref{fig:dustmass}, where the lenticulars are shown to have the same mean specific dust mass (-3.35) as the ellipticals (-3.36). However, a KS-test of the stellar mass distributions show them to be marginally different at the 1$\%$ level (KS-prob=0.009). The straight-line fit to the ellipticals suggests a slightly stronger trend for specific dust mass to change with stellar mass (r$_{p}$=-0.71) than the trend for lenticulars (r$_{p}$=-0.41). Smaller number densities for the ellipticals weakens this trend somewhat. This result nevertheless indicates that the two different types of ETGs have similar dust properties and therefore that there is some similarity in the way dust in these galaxies evolves and consequently, in the evolution of the galaxy itself.

%This result is strongly in support of downsizing theories. Some of our \textit{SubS} ETGs may not yet have built up their stellar masses through interactions with other galaxies, and therefore much of their cold dust has not been destroyed during these interactions.

We explore the separate morphological properties further in Figs. \ref{fig:multi1} and \ref{fig:multi2} by plotting different parameters against the absolute r band magnitudes and normalised dust masses respectively. In Figs. \ref{fig:multi1}(a) and (b) we show how both normalised dust mass and total dust mass vary with absolute magnitude. For both galaxy types we see a clear trend emerging here, which is similar to the trend identified earlier in $\S$\ref{sec:dustydust}. This is expected since magnitude and stellar mass are related (see discussion in $\S$\ref{sec:dustydust}). These plots show that the brightest ETGs contain proportionally the least dust and vice versa (Fig. \ref{fig:multi1}(d)). Once again, by breaking ETGs down to elliptical and lenticular, we explore their comparative properties and find the brightest ETGs are elliptical, with low specific dust mass (Fig. \ref{fig:multi1}(d)) and high dust mass (Fig. \ref{fig:multi1}(e)). On the faint end, we see a well mixed distribution of both types of ETGs. This, and a KS-test, indicates no preference for lenticulars to have different specific dust masses than ellipticals.

The wide distribution of dust temperatures arising from our modified Planck function fits is also highlighted in Fig. \ref{fig:multi1}(c) and (f). Our ETGs extend to very low dust temperatures, with errors of typically $\pm$2 K. This could be attributed to the fact that this is a very large sample of ETGs observed by very sensitive detectors and the galaxies may contain dust temperatures which could not physically have been measured up until now. However, it should be noted that the galaxy with the lowest temperature ($\sim$9 K) has large PACS measurement errors.

All the other ETGs in our sample have dust temperatures higher than 10 K. The distributions of ellipticals versus lenticulars look quite similar, although the lowest temperatures ($<$15 K) belong predominantly to ellipticals. KS-testing shows the distributions are not significantly different. Based on this result, it appears that lenticulars not only contain similar (specific) dust masses to ellipticals, but also similar dust temperatures. The lack of observable trend of temperature with luminosity also implies that there is no direct correlation between distributions of dust temperature and stellar mass for the two types of ETGs.

The distributions of environmental, S\'ersic and colour properties of the galaxies as a function of normalised dust mass are examined in Fig. \ref{fig:multi2}. Fig. \ref{fig:multi2}(a) indicates that there is no significant difference in the distribution of surface densities for the two morphologies, which is confirmed by a KS-test. Its counterpart scatter plot (Fig. \ref{fig:multi2}(d)) shows no observable trend for these properties, although the few ellipticals with an intermediate range of specific dust mass seem to be clustered towards the lower densities.

%However, its counterpart scatter plot (Fig. \ref{fig:multi2}(d)) indicates that the elliptical galaxies are non-existent at densities of $\gtrsim$7 galaxies Mpc$^{-2}$ and the lowest density regions only contain ETGs with high specific dust masses. This makes a case for ETGs with little dust residing in denser environments, supporting downsizing theories where galaxies in the centres of clusters evolve and are stripped of their dust and gas more quickly than those galaxies in field environments.

Figs. \ref{fig:multi2}(b) and (e) show the S\'ersic (n$_{r}$) distributions of these morphologies and highlight a useful characteristic of ellipticals. Although there are only small numbers of ellipticals, approximately a third of them are at lower S\'ersic (n$\sim$1). However, Fig. \ref{fig:multi2}(e) also indicates that on average, these low S\'ersic ETGs are those with the highest specific dust masses (r$_{p}$=0.1). This again supports the idea that dust is causing the lowering of S\'ersic index, as predicted by \citet{pastrav_2013}.

Finally we examine UV-optical colour in Figs. \ref{fig:multi2}(c) and (f). Although this trend was seen previously in Fig. \ref{fig:NUVdust}, in this plot we see the significance of different morphologies. Once again we find different groupings of ellipticals, unlike the solidly filled range presented by the lenticulars. A KS-test shows NUV-r distributions to be significantly different at well below the 1$\%$ level (KS-prob=0.0009). Ellipticals are grouped either as very blue and dusty, or red with a range of specific dust masses. Such a distribution of specific dust mass at the red end indicates that dust may be contributing to reddening in this plot, otherwise we would expect galaxies with such high levels of dust to occupy the bluer end of this distribution.

%Re:lensing: shouldn't have to worry about it at such low redshift.
\section{Summary and Conclusions}\label{sec:conc}

This work is the first in a two-part series exploring  intrinsic and dust properties of early-type galaxies. By using visual classifications applied to a low-z sample of GAMA galaxies from \citet{kelvin_2013}, we created two ETG samples from the GAMA data. These are the $SubS$ - galaxies classified as ETGs with H-ATLAS SPIRE 5$\sigma$ detections - and the $OptS$ - galaxies classified as ETGs without H-ATLAS SPIRE detections.

We explored a series of optical and UV properties for these two samples. These included concentration and S\'ersic index, stellar masses, luminosities, optical colours, and effective radii. GALEX NUV detections were used to explore UV-optical colours of our galaxies and the environments of the ETGs were examined by way of 5$^{th}$ nearest neighbour densities and group membership data. The results for these tests are summarised here:

\begin{enumerate}[label=\roman{*})]

\item{We find a 29$\%$ H-ATLAS detection rate for ETGs in the equatorial field. The detection rate for pure sub-mm detected ellipticals is 10$\%$ (see $\S$\ref{sec:samples}).}
\item{H-ATLAS detected ($SubS$) ETGs are shown to have lower concentrations and S\'ersic indices than undetected ($OptS$) ETGs. It has been shown that the presence of dust can lower S\'ersic index \citep{pastrav_2013}, an effect that we consider to be a possible explanation for our findings. This effect will be further explored in Paper II of this series, where we will investigate whether this is true for our samples in all wavebands and at higher redshifts. }
\item{H-ATLAS detected ETGs are shown to be typically optically brighter, bluer and larger than undetected ETGs. However, differences in the stellar mass distributions of the samples are found to be not significant.}
\item{Our H-ATLAS detected sample shows significantly brighter NUV luminosities and bluer colours in the UV-optical than the undetected sample. This colour trend is in agreement with previous works such as \citet{Rowlands_2011} and \citet{dariush_2011}. However, the H-ATLAS detected ETGs are redder than LTGs from the same parent sample, and therefore dominate an intermediate NUV-r colour space.}
\item{The H-ATLAS detected ETGs are shown to inhabit sparser environments, particularly for the fainter ETG subset. By comparing our ETGs in both small and large galaxy groups with ungrouped ETGs, we find that for the $SubS$, higher mass stellar systems are typically associated with larger group sizes and denser environments.}

\end{enumerate}

To investigate dust properties, we fitted modified single-temperature Planck functions to the 250$\mu$m 5$\sigma$ detected $SubS$ galaxies containing $\ge$3$\sigma$ emission in the SPIRE 350$\mu$m waveband. We then presented characteristics of the $SubS$ based on resultant best-fit temperatures and associated dust masses, and explored trends with optical and UV properties. These results can be summarised as follows:

\begin{enumerate}[label=\roman{*})]
\item{ETGs in our H-ATLAS detected sample are shown to contain dust masses ranging from 8.1$\times$10$^{5}$-3.5$\times$10$^{8}$M$_{\odot}$, with a range of rest-frame temperatures from 9-30K. These dust masses are consistent with previous $Herschel$ work such as \citet{skibba_2011}, but lower than results from previous H-ATLAS work in \citet{Rowlands_2011}. These results may differ due to our larger sample size and our lower redshift limit.}
\item{We discover a strong trend for specific dust mass to decrease with redder (NUV-r) colour. This implies that the dustiest ETGs have the bluest colours, linking recent star formation with a heavier dust presence.}
\item{We show that the faintest H-ATLAS detected ETGs with the lowest stellar masses have higher specific dust masses and lower S\'ersic indices compared to the brightest H-ATLAS detected ETGs. This result is consistent with downsizing, as the most massive, brightest galaxies are more similar to the $OptS$ ETGs in terms of their specific dust masses and S\'ersic profiles.}
\item{By splitting our dusty ETG sample into separate morphologies of elliptical and lenticular galaxies, we gauge how they contribute to the measured dust properties. We find no difference in the dust masses (whole or specific) or temperatures of both ETG types. However, we find significant differences in their NUV-r colour and stellar mass distributions.}
\item{We show that ellipticals may be grouped into two sets: the faint, blue, low mass, relatively dusty ellipticals and bright, red, massive, ellipticals with lower specific dust masses. This leads us to the conclusion that we may be studying two different populations of ellipticals, or at least two different age ranges.}
\end{enumerate}

We intend to further explore these separate populations of ETGs by pushing our samples out to higher redshifts in Paper II of this series. As we can no longer rely on visual classifications at these redshifts, we will explore proxies for morphology. We will then repeat our study of ETG optical and dust characteristics, and compare them with these low redshift samples.

%KEEP THIS PARAGRAPH FOR LINKING SECTIONS IN LE THESIS The ETG samples in this series of papers have been selected and analysed to set up future work in examining the \textit{SubS}'s cold dust properties. The isothermal SED fitting described above in $\S$ \ref{sec:sec4} outputs some basic properties for the sub-mm selected galaxies, but doesn't tell us much about which mechanism powers the dust emission within their galaxies' interstellar media. We intend to discover more about the distribution and physical properties of dust in ETG ISM. Such properties include the grain temperatures and overall dust masses, but also the physical mechanisms which heat galactic dust. This work will begin with fitting the MIR/FIR/sub-mm waveband data with several template models. The first model will account for the more well-known radiative heating by photons in the ISM, whereas the second model will fit the data with parameters based on collisional heating. Our primary aims are to develop an understanding of what the main source of dust heating is within sub-mm emitting ETGs.

\section*{Acknowledgments}

We would like to thank the anonymous referee whose comments led to important improvements in this paper. Additionally, we would like to thank Giovanni Natale for his comments and advice in the writing and proofing of this paper. NKA acknowledges the support of the Science and Technology Facilities Council. The \textit{Herschel}-ATLAS is a project with \textit{Herschel}; which is an ESA space observatory with science instruments provided by European-led Principal Investigator consortia and with important participation from NASA. The H-ATLAS website is http://www.h-atlas.org/. GAMA is a joint European-Australasian project based around a spectroscopic campaign using the Anglo-Australian Telescope. The GAMA input catalogue is based on data taken from the Sloan Digital Sky Survey and UKIRT Infrared Deep Sky Survey. Complementary imaging of the GAMA regions is being obtained by a number of independent survey programs including GALEX MIS, VST KIDS, VISTA VIKING, WISE, \textit{Herschel}-ATLAS, GMRT and ASKAP providing UV to radio coverage. GAMA is funded by the STFC (UK), the ARC (Australia), the AAO, and the participating institutions. The GAMA website is http://www.gama-survey.org/.

\bibliographystyle{mn2e}
	\bibliography{diagbib}

\end{document}